\documentclass[aps,prr,reprint,superscriptaddress]{revtex4-1}
\usepackage{lmodern}
\usepackage{graphicx}
\usepackage{braket}
\usepackage{amsfonts, amsmath, amsthm, amssymb}
\usepackage{bbold}
\usepackage{subfigure}
\usepackage{hyperref}
\hypersetup{colorlinks=true,linkcolor=blue,citecolor=blue,urlcolor=blue}
\def\equationautorefname~#1\null{(#1)\null}
\usepackage{cleveref}
\usepackage[utf8]{inputenc}
\usepackage[T1]{fontenc}
\Crefname{figure}{Fig.}{Figs.}
\newcommand*{\eqnref}[1]{%
\begingroup
Eq. \autoref{#1}%
\endgroup}
\makeatletter
\newsavebox{\@brx}
\newcommand{\llangle}[1][]{\savebox{\@brx}{\(\m@th{#1\langle}\)}%
  \mathopen{\copy\@brx\kern-0.5\wd\@brx\usebox{\@brx}}}
\newcommand{\rrangle}[1][]{\savebox{\@brx}{\(\m@th{#1\rangle}\)}%
  \mathclose{\copy\@brx\kern-0.5\wd\@brx\usebox{\@brx}}}
\makeatother

\begin{document}

\title{Optical imprinting of superlattices in two-dimensional materials}

\author{Hwanmun Kim}
\altaffiliation{Contributed equally}
\affiliation{Joint Quantum Institute, NIST/University of Maryland, College Park, Maryland 20742, USA}
\affiliation{Department of Physics, University of Maryland, College Park, Maryland 20742, USA}

\author{Hossein Dehghani}
\altaffiliation{Contributed equally}
\affiliation{Joint Quantum Institute, NIST/University of Maryland, College Park, Maryland 20742, USA}
\affiliation{Departments of Electrical and Computer Engineering and Institute for Research in Electronics and Applied Physics, University of Maryland, College Park, Maryland 20742, USA}

\author{Hideo Aoki}
\affiliation{Department of Physics, The University of Tokyo, Hongo, Tokyo 113-0033, Japan}
\affiliation{National Institute of Advanced Industrial Science and Technology (AIST), Tsukuba, Ibaraki 305-8568, Japan}
\author{Ivar Martin}
\affiliation{Materials Science Division, Argonne National Laboratory, Argonne, Illinois 60439, USA}
\author{Mohammad Hafezi}
\affiliation{Joint Quantum Institute, NIST and University of Maryland, College Park, Maryland 20742, USA}
\affiliation{Department of Physics, University of Maryland, College Park, Maryland 20742, USA}
\affiliation{Departments of Electrical and Computer Engineering and Institute for Research in Electronics and Applied Physics, University of Maryland, College Park, Maryland 20742, USA}

\date{\today}

\begin{abstract}
We propose an optical method of shining circularly polarized and spatially periodic laser fields to imprint superlattice structures in two-dimensional electronic systems. By changing the configuration of the optical field, we synthesize various lattice structures with different spatial symmetry, periodicity, and strength. We find that the wide optical tunability allows one to tune different properties of the effective band structure, including Chern number, energy bandwidths, and band gaps. The \textit{in situ} tunability of the superlattice gives rise to unique physics ranging from the topological transitions to the creation of the flat bands through the kagome superlattice, which can allow a realization of strongly correlated phenomena in Floquet systems. We consider the high-frequency regime where the electronic system can remain in the quasiequilibrium phase for an extended amount of time. The spatiotemporal reconfigurability of the present scheme opens up possibilities to control light-matter interaction to generate novel electronic states and optoelectronic devices.
\end{abstract}

\maketitle
\section{Introduction}
A superlattice structure in two-dimensional (2D) materials has opened a new way to engineer electronic bands, starting with the investigation on a honeycomb superlattice structure in monolayer graphene \cite{PhysRevLett.71.4389}.  Recently, Moir\'e pattern in a twisted-bilayer van der Waals heterostructure has been immensely successful in generating a variety of band structures, including Hofstadter butterfly \cite{PhysRevB.84.035440,spanton2018observation} and flat bands \cite{bistritzer2011moire,PhysRevB.86.155449,PhysRevLett.122.106405,cao2018unconventional,cao2018correlated}. These bands can induce intriguing strongly correlated phases such as fractional Chern insulator \cite{spanton2018observation}, anomalous Hall phase \cite{sharpe2019emergent,PhysRevX.9.031021}, Mott insulating phase \cite{cao2018correlated,choi2019electronic,chen2019evidence,PhysRevX.8.031089}, nontrivial magnetic phases \cite{PhysRevLett.119.107201,PhysRevB.98.075109,sharpe2019emergent,PhysRevLett.120.266402}, and superconductivity \cite{cao2018unconventional,yankowitz2019tuning,PhysRevX.8.031089,PhysRevLett.121.257001,PhysRevLett.122.257002}. Yet, this passive way of creating a superlattice has been largely limited by the microscopic structure of the 2D materials since different samples should be prepared for different superlattice structures. Therefore, it is interesting to find alternative ways to synthesize a spatiotemporal structure in 2D materials.

At the same time, the recent progress in the beam-shaping technique has enabled the generation of arbitrary beam patterns with high resolution comparable to the optical wavelengths \cite{zupancic2016ultra,barredo2016atom,endres2016atom,barredo2018synthetic,schine2019electromagnetic,fazal2011optical}, which already found remarkable successes in ultracold-atom systems \cite{tai2017microscopy,lukin2019probing,chiu2019string,PhysRevLett.122.173201,PhysRevX.9.041052}. This wide tunability of light can be naturally applied to 2D electronic systems to imprint arbitrary superlattices, regardless of the underlying microscopic lattice structure. This is particularly interesting in the context of the ``Floquet topological insulator,'' where the illumination of circularly polarized (CP) light can turn a trivial system into a topological insulator \cite{PhysRevB.79.081406,kitagawa2011transport,lindner2011floquet,wang2013observation, mciver2019light,PhysRevB.90.115423,PhysRevResearch.1.023031,li2019floquet,katz2019optically,PhysRevLett.107.276601}.  

\begin{figure}[t]
\centering
\includegraphics[width=\linewidth]{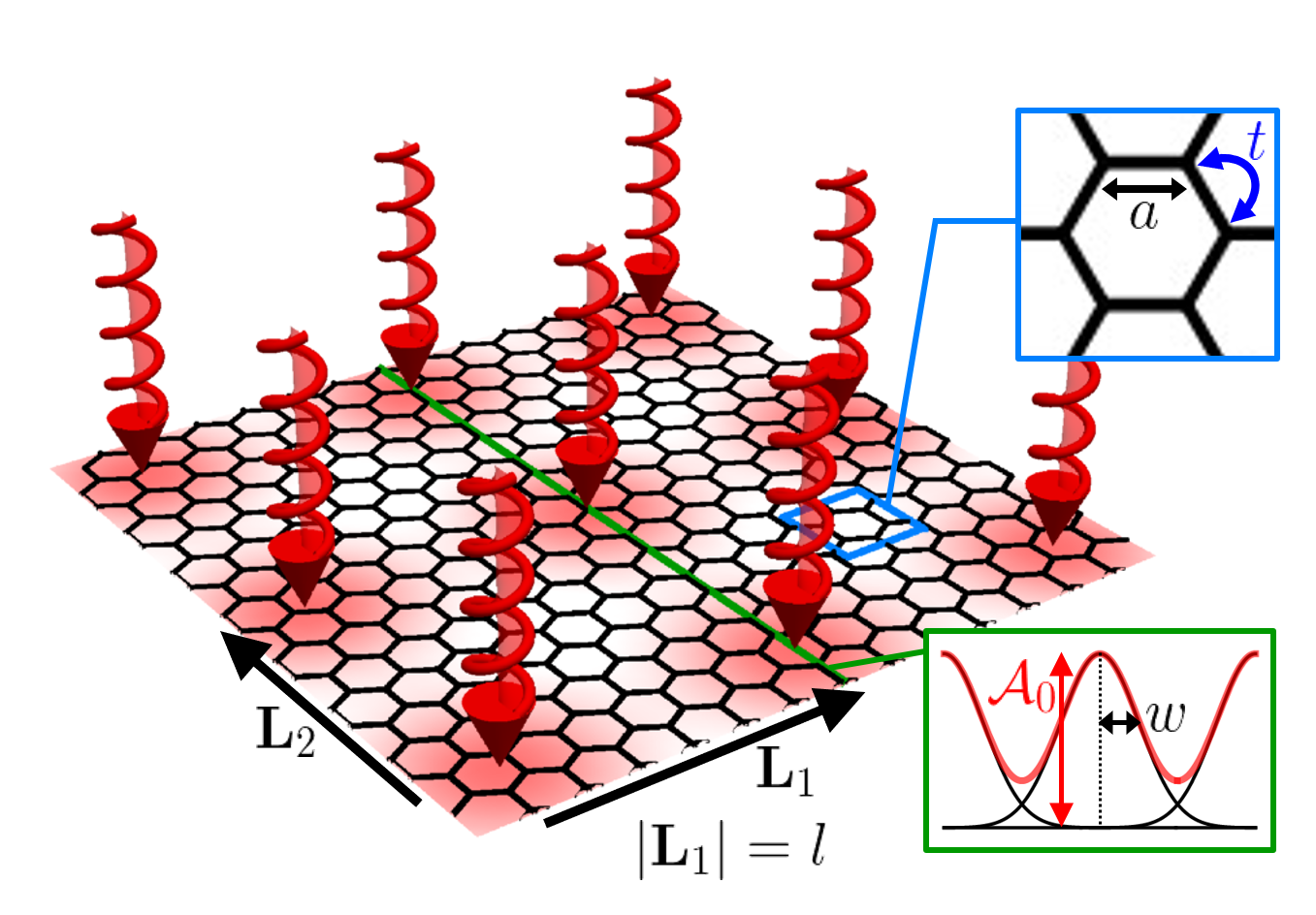}\subfigure{\label{F1a}}\subfigure{\label{F1b}}
\caption{A 2D material irradiated by a spatially periodic CP light with frequency $\omega$. Here, we use the example of a monolayer graphene. The superposition of multiple CP Gaussian beams generates a periodic amplitude pattern $A_0(\mathbf{r})$ with translation vectors $\mathbf{L}_1$ and $\mathbf{L}_2$. $|\mathbf{L}_1|=l$. Upper inset: We denote the interatomic distance of the graphene as $a$ and the tight-binding energy between the nearest neighbors as $t$. Lower inset: Each Gaussian beam has a peak amplitude $\mathcal{A}_0$ and a half waist $w$ (black lines). The overall beam amplitude (red line) results from the superposition of the Gaussian beams.}\label{F1}
\end{figure}

In this paper, we propose a method to create superlattice structures in a 2D material by shining spatially periodic laser beams, as schematically shown in \Cref{F1}. We illustrate the idea with an example of monolayer graphene irradiated by a circularly polarized beam with a superlattice structure, where the beam amplitude is spatially periodic. To demonstrate the tunability of this superlattice structure and unique physics originating from the superlattice, we first study the case of a square superlattice and explore the topological phase transition induced by varying the superlattice size. Then, we investigate the topological phase transitions, when the square superlattice is sheared to a stretched hexagonal one. In particular, we examine the relationship between this topological phase transition and the role of lattice geometry in creating complex tunneling phases. Further, we demonstrate the possibility of creating more exotic lattices by superposing multiple lattices, with an example of tuning between a hexagonal and a kagome lattice where the flat bands can be obtained. These flat bands particularly can harbor strongly correlated phenomena in Floquet systems.   

\section{Graphene with spatially patterned light}
Let us consider a monolayer graphene with the inter-atomic distance $a$ and the tight-binding energy $t$ between the nearest neighbors. The low-energy description for this monolayer graphene under the electromagnetic field $\mathbf{A}(\mathbf{r},t)$ is given by
\begin{eqnarray}\label{H_coupling}
H = v\left[ \mathbf{p}+e\mathbf{A}(\mathbf{r},t)\right] \cdot \left( \tau_z\sigma_x\mathbf{\hat{x}}+\sigma_y\mathbf{\hat{y}} \right),
\end{eqnarray}
where $\sigma_x,\sigma_y,\sigma_z$ are Pauli matrices acting on sublattice degrees of freedom, $v=(3/2)ta$ is the Fermi velocity at Dirac points, and $\tau_z=\pm 1$ is the valley index \cite{RevModPhys.81.109}. In particular, if we shine the CP beam with spatial amplitude pattern $\mathbf{A}(\mathbf{r},t)=A_0(\mathbf{r})e^{i\omega t}(\mathbf{\hat{x}}+i\mathbf{\hat{y}})+\text{c.c.}$ (\Cref{F1}), the effective Floquet Hamiltonian to the first order in $\omega^{-1}$ becomes \cite{PhysRevB.79.081406,PhysRevLett.110.200403,PhysRevX.4.031027,HDehghani2014Dissipative,[{In a Floquet system, the quantum Hall physics is dominated by the non-equilibrium distribution that depends on whether the system is isolated or coupled to a reservoir where photoexcitation competes with bath-induced cooling, as elaborated in }]HDehghani2015Out,*PhysRevB.99.014307,eckardt2015high,PhysRevB.93.144307}
\begin{eqnarray}\label{eqn_Heff}
H_\text{eff} = v(\tau_z p_x\sigma_x + p_y\sigma_y) + \tau_z\frac{4e^2 v^2}{\omega}\left|A_0(\mathbf{r})\right|^2 \sigma_z.
\end{eqnarray}
We denote the peak amplitude of $A_0(\mathbf{r})$ as $\mathcal{A}_0$. Then, \eqnref{eqn_Heff} becomes a valid description when frequency $\omega$ is high enough ($\omega \gg ev\mathcal{A}_0$) and the amplitude varies in length scale larger than $a$ ($\mathcal{A}_0/\text{max}\left\lbrace |\nabla A_0(\mathbf{r})| \right\rbrace \gg a$). For brevity, we set $\hbar=1$ from here on.

We specifically study the superlattice structure created by a spatially periodic amplitude $|A_0(\mathbf{r})|=|A_0(\mathbf{r}+\mathbf{L}_1)|=|A_0(\mathbf{r}+\mathbf{L}_2)|$. While the 2D material with spatioally modulated beams has been studied in the different contexts \cite{PhysRevLett.110.016802,PhysRevB.88.224106,morina2018optical}, here we investigate the generation of a superalttice with spatially periodic beams. In particular, to make the beam experimentally relevant, we consider the superposition of CP Gaussian beams positioned on the superlattice,
\begin{eqnarray}
A_0(\mathbf{r}) = \sum_{n_1,n_2} \mathcal{A}_0 \exp\left( -\frac{|\mathbf{r}-n_1\mathbf{L}_1-n_2\mathbf{L}_2|^2}{2 w^2} \right),
\end{eqnarray}
where $w$ is the radius of each Gaussian beam. This beam configuration is achievable with recent progress in beam-shaping technologies \cite{zupancic2016ultra,barredo2016atom,endres2016atom,barredo2018synthetic,schine2019electromagnetic,fazal2011optical}. For the cases $|\mathbf{L}_1|,|\mathbf{L}_2|=l\gg a$, the Brillouin-zone folding occurs on a momentum scale $1/l$. Furthermore, the hybridization of Floquet sidebands is suppressed  for $v/l\ll\omega$ so that the low-energy description is captured by \eqnref{eqn_Heff} (see Appendix A). We obtain Bloch eigenstates $\ket{\psi_{m,\mathbf{k}}}$ and eigenenergies $ E_{m,\mathbf{k}}$, where $m$ is the band index and $\mathbf{k}$ is the crystal momentum within the Brillouin zone set by reciprocal lattice vectors of $\mathbf{L}_1$ and $\mathbf{L}_2$. Note that \eqnref{eqn_Heff} preserves particle-hole symmetry ($\sigma_x H_{\text{eff}}^* \sigma_x = -H_{\text{eff}}$) and therefore the energy spectrum is symmetric with respect to the zero energy. Also, $\sigma_y H_{\text{eff}} \sigma_y = \left. H_{\text{eff}}\right|_{\tau_z\to -\tau_z}$, so two valleys have the same spectrum and eigenstates up to a unitary operation, $\sigma_y$. This also ensures that both valleys have the same Chern number. For brevity, let us only consider the $\tau_z=1$ valley from now on.

\section{Illumination of square superlattice}
We first consider the simplest case of a square superlattice, $\mathbf{L}_1=l\mathbf{\hat{x}}$ and $\mathbf{L}_2=l\mathbf{\hat{y}}$. Before directly diagonalizing \eqnref{eqn_Heff}, we can make some speculations. First of all, the contribution from the spatial average of $|A_0(\mathbf{r})|$ opens up the gap around the zero energy ($\Delta_b$) as in the case of the graphene under the CP uniform light, where the Chern number, $\mathcal{C}_1$, of the first band above $E=0$ is nonzero \cite{PhysRevB.79.081406,kitagawa2011transport,PhysRevLett.110.016802,lindner2011floquet,PhysRevX.3.031005}. $\mathcal{C}_1$ remains nonzero for small $l$, as far as the maximum kinetic energy within the Brillouin zone, which is of the order of $v/l$, is much larger than the spatial Fourier components of the $\sigma_z$ term in \eqnref{eqn_Heff}, which is of the order of $e^2v^2\mathcal{A}_0^2/\omega$. On the other hand, as $l\to\infty$, the contribution of the kinetic term becomes negligible and therefore the bands become flat. Also, the Bloch wavefunctions look similar regardless of $\mathbf{k}$ and therefore the bands become topologically trivial. Therefore, there must be a topological phase transition where $\mathcal{C}_1$ changes from a nonzero value to zero as we increase $l$. This topological transition would occur at a superlattice size that makes the two energy scales $e^2v^2\mathcal{A}_0^2/\omega$ and $v/l$ comparable to each other. For a succinct description of this phase transition, we use the rescaled superlattice size
\begin{eqnarray}
\chi = (v e^2 \mathcal{A}_0^2 /\omega)l
\end{eqnarray}
so that the critical superlattice size $\chi_c$ is $O(1)$. Here, $\chi$ represents the ratio of the effective superlattice potential over the kinetic energy.

\begin{figure}[t]
\centering
\includegraphics[width=\linewidth]{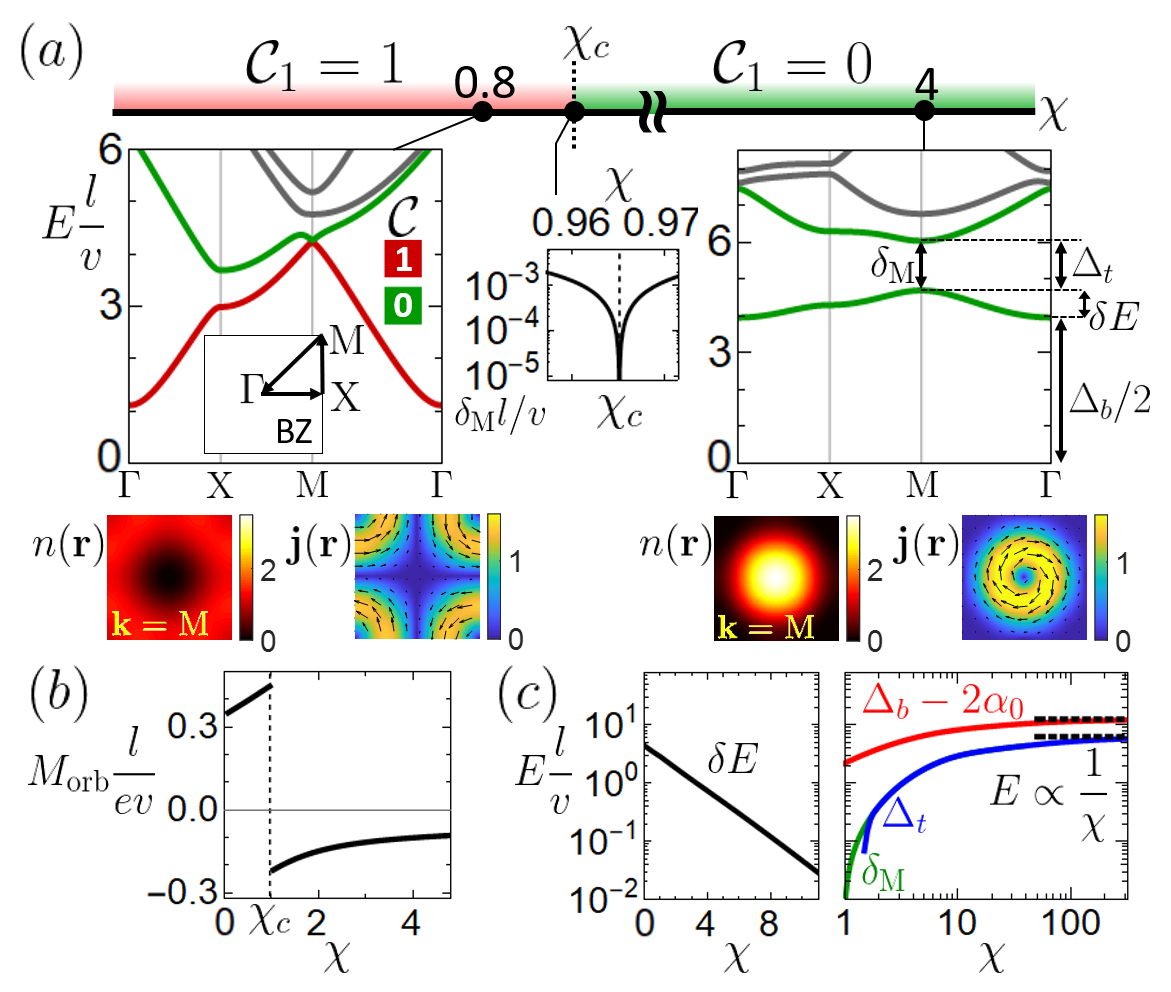}\subfigure{\label{F2a}}\subfigure{\label{F2b}}\subfigure{\label{F2c}}
\caption{(a) Energy spectrum for square superlattices with different superlattice size $\chi$. We set $\mathcal{A}_0=0.006(ea)^{-1}$, $\omega=0.06t$, and $w/l=0.3$. Only the positive-energy spectrum is shown for simplicity. The Chern numbers of low-lying bands, $\mathcal{C}$, are presented as colors. The topological phase transition occurs at $\chi_c=0.965$. Upper inset: Direct gap at $\mathbf{k}=\text{M}$ between the first- and the second-lowest positive band ($\delta_\text{M}$) is plotted in the vicinity of $\chi_c$. Lower insets: The particle density $n(\mathbf{r})$ and current density $\mathbf{j}(\mathbf{r})$ of the Bloch wavefunction of the lowest positive band at $\mathbf{k}=\text{M}$ are shown for $\chi=0.8<\chi_c$ and $\chi=4>\chi_c$. In the density plots, the centers of the Gaussian beams are located at the corners of the plotted region. The particle density is shown in units of $l^{-2}$. The amplitude of the current density is presented with the color in units of $ev/l^2$ and the direction of $\mathbf{j}(\mathbf{r})$ is represented by arrows. (b) Orbital magnetization $M_\text{orb}$ for the lowest positive band for different superlattice sizes. (c) For the lowest positive band, we plot the energy gap below the band ($\Delta_b$), the energy gap above the band ($\Delta_t$), the direct band gap at $\mathbf{k}=\text{M}$ ($\delta_M$), and the bandwidth ($\delta E$) with respect to the superlattice size $\chi$. $\alpha_0$ is the minimum value of $(4e^2 v^2/\omega)|A_0(\mathbf{r})|^2$. The black dashed lines are asymptotic lines showing that $El/v$ is constant, indicating $E\propto\chi^{-1}$. }\label{F2}
\end{figure}

To study the detail of this topological phase transition, we numerically diagonalize \eqnref{eqn_Heff} as shown in \Cref{F2a}. Along with the energy spectrum, we present the Chern number $\mathcal{C}$ of each band calculated based on Ref. \cite{fukui2005chern}. In \Cref{F2}, we set $\mathcal{A}_0=0.006(ea)^{-1}$, $\omega=0.06t$, and $w/l=0.3$. With these parameters, we can check that the topological phase transition occurs at $\chi_c=0.965$, which is close to 1. This topological transition accompanies the direct gap closing at $\mathbf{k}=\text{M}$ and the band inversion between the first- and second-lowest positive-energy bands. To see this, we compare the particle and current densities of the lowest positive-energy band's wave function at the direct gap closing point. Here, for the Bloch wavefunction of the $m$th band, $\psi(\mathbf{r})=\braket{\mathbf{r}|\psi_{m,\mathbf{k}}}$, the particle and current densities are given by
\begin{eqnarray}
n(\mathbf{r}) &=& \psi^\dag(\mathbf{r}) \psi(\mathbf{r}), \\
\mathbf{j}(\mathbf{r}) &=& -e\psi^\dag(\mathbf{r}) \frac{\partial H_\text{eff}}{\partial \mathbf{p}} \psi(\mathbf{r})
= -ev\psi^\dag(\mathbf{r})\left( \sigma_x\mathbf{\hat{x}} + \sigma_y\mathbf{\hat{y}} \right) \psi(\mathbf{r}). \nonumber
\end{eqnarray}
The comparison of $n(\mathbf{r})$ and $\mathbf{j}(\mathbf{r})$ before ($\chi=0.8$) and after ($\chi=4$) the transition point shows a drastic change in the wave function, which signifies that the band inversion has occurred in the phase transition. In the current density plot, one can also find that the circulation direction of the electron flips as the band inversion occurs. This phenomenon can also be captured in the calculation of the $m$th band contribution to the orbital magnetization \cite{PhysRevLett.95.137205,PhysRevLett.95.137204,PhysRevLett.99.197202},
\begin{eqnarray}\label{eqn_Morb}
M_\text{orb}= \text{Im}\int \frac{d^2\mathbf{k}}{(2\pi)^2}
e\frac{\partial\bra{u_{m,\mathbf{k}}}}{\partial k_x}
\left( H_{\mathbf{k}} + E_{m,\mathbf{k}} \right)
\frac{\partial\ket{u_{m,\mathbf{k}}}}{\partial k_y}, \quad
\end{eqnarray}
where $\ket{u_{m,\mathbf{k}}}=e^{-i\mathbf{k}\cdot\mathbf{r}}\ket{\psi_{m,\mathbf{k}}}$ and $H_{\mathbf{k}}=e^{-i\mathbf{k}\cdot\mathbf{r}} H_{\text{eff}} e^{i\mathbf{k}\cdot\mathbf{r}}$. In \Cref{F2b}, one can see that $M_\text{orb}$ of the lowest positive band shows the sign flip at the phase transition point, agreeing with the observation in the current density plots.
We also remark that even if this topological phase transition theoretically exists regardless of the Gaussian beam size, it is desirable to keep $w$ comparable to $l$ for experimental realizations since a fainter superlattice will imply a smaller direct band gap.

This topological phase transition could be experimentally detected in several ways. The change in $\mathcal{C}_1$ causes the difference in the Hall current carried by the chiral edge state, and such difference can be revealed by transport measurements, similar to  Ref. \cite{mciver2019light}. For the bulk property, one can measure the orbital magnetization, where the sudden jump would be observed at the phase transition shown in \Cref{F2b}.

As the superlattice size $\chi$ increases, the electrons become localized at the local minima of $|A_0(\mathbf{r})|$. This provides an explanation for the exponential suppression of the bandwidth of the lowest positive-energy band ($\delta E$) in $\chi$ [\Cref{F2c}]. For well-localized electrons, the dynamics can effectively be described by a tight-binding model, and the tunneling energy of that model is approximately given by the WKB integrals. This integral decays exponentially with the distance between the superlattice sites, so the bandwidth decreases exponentially as well. The band gaps ($\Delta_b$, $\Delta_t$, $\delta_M$) decay as $O(\chi^{-1})$, where the details of this band gap scaling are explained in the Appendix B.

\section{Superlattice shearing}

\begin{figure}[t]
\centering
\includegraphics[width=\linewidth]{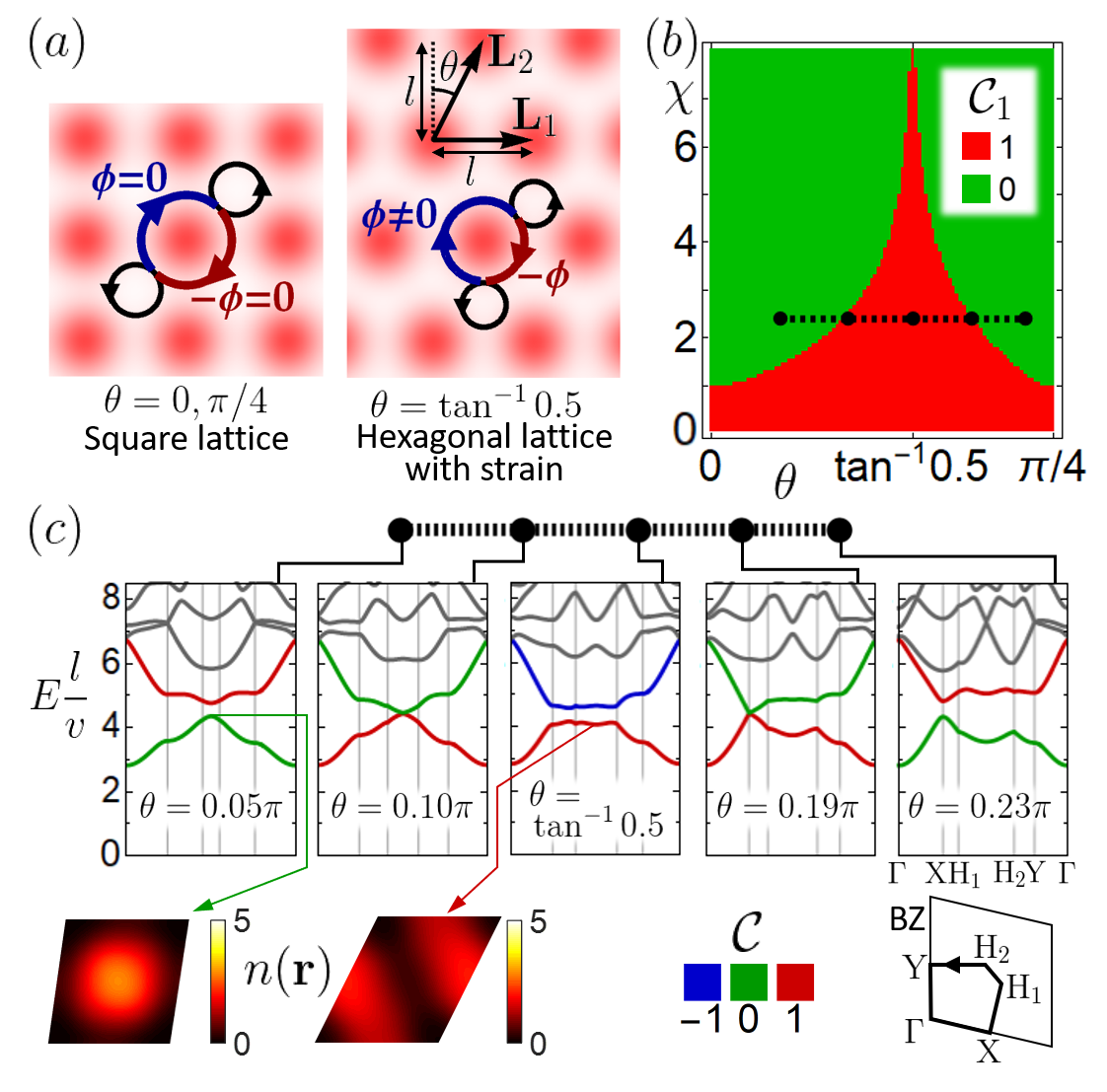}\subfigure{\label{F3a}}\subfigure{\label{F3b}}\subfigure{\label{F3c}}\subfigure{\label{F3d}}
\caption{(a) We shear a square lattice by angle $\theta$. Tunneling between two sites can be understood as the flow of chiral edge currents around each Gaussian CP beam. If the system has reflection symmetry around the line connecting the two sites, this tunneling should be real. Otherwise, the tunneling can have a complex phase. As examples, the next-nearest-neighbor tunnelings for the $\theta=0,\pi/4$ case and the $\theta=\tan^{-1}(1/2)$ case are presented. (b) The Chern number of the lowest positive energy band $\mathcal{C}_1$ is shown as a phase diagram between the shearing angle $\theta$ and the superlattice size $\chi$. (c) Energy spectra for $\chi=2.4$ at selected angles are shown where the colors of low-lying bands represent the Chern numbers. The particle density in units of $l^{-2}$ is plotted for angles before and after the phase transition.}\label{F3}
\end{figure}

To further investigate the role of the superlattice geometry, let us shear the square superlattice by angle $\theta$ so that $\mathbf{L}_1=l\mathbf{\hat{x}}$ and $\mathbf{L}_2=l(\tan\theta\mathbf{\hat{x}}+\mathbf{\hat{y}})$. From the perspective of the Floquet Chern insulator created by uniform CP light, in a large superlattice size limit where the tight-binding description is valid, we might interpret the electron tunneling between superlattice sites as the chiral currents around the strongly irradiated region. That is, the paths that these chiral currents flow would give the major contribution to the path integral from one superlattice site to another. In this viewpoint, two superlattice sites can have a \textit{complex} tunneling phase between them if the system has no reflection symmetry along the line connecting the two sites [\Cref{F3a}], which is analogous to Ref. \cite{hafezi2011robust}. Then we can see that the tunneling terms of the tight-binding model for the square lattice ($\theta=0$ and $\theta=\pi/4$) are real. At angles close to $\theta=\tan^{-1}(1/2)$, the localized electrons form a hexagonal superlattice under a uniform strain and can have complex tunneling phases between the next-nearest neighbors. Then we can construct a tight-binding model for the lowest positive band similar to the Haldane model \cite{PhysRevLett.61.2015}, as explained in the Appendix C. Similar to the Haldane model, a complex tunneling phase in the next-nearest-neighbor tunneling makes $\mathcal{C}_1$ nonzero at this angle. With these considerations, we can predict successive topological phase transitions as we increase $\theta$ from 0 to $\pi/4$.

We obtain the phase diagram numerically in \Cref{F3b} by calculating the Chern number of the lowest positive-energy band for each value of $\chi$ and $\theta$. As we predicted, we can observe the successive topological phase transitions at $\chi$ larger than a certain value, which corresponds to the phase transition point described in \Cref{F2}. Another salient feature is that the $\mathcal{C}_1=1$ regime very sharply blows up toward the angle $\theta=\tan^{-1}(1/2)$, at which the $\chi$ region for $\mathcal{C}_1=1$ diverges. This can be explained by combining the fact that the size of tunneling strengths decreases exponentially with the distance between the superlattice points and another fact that the Dirac cones can disappear and the topologically trivial gap opens in the extreme strain (see Appendix C). We can also see that the topological phase transition also accompanies the gap closing and the band inversion, as shown in the particle density plots [\Cref{F3c}].

\section{Hexagonal lattice to kagome lattice}

\begin{figure}[h]
\centering
\includegraphics[width=\linewidth]{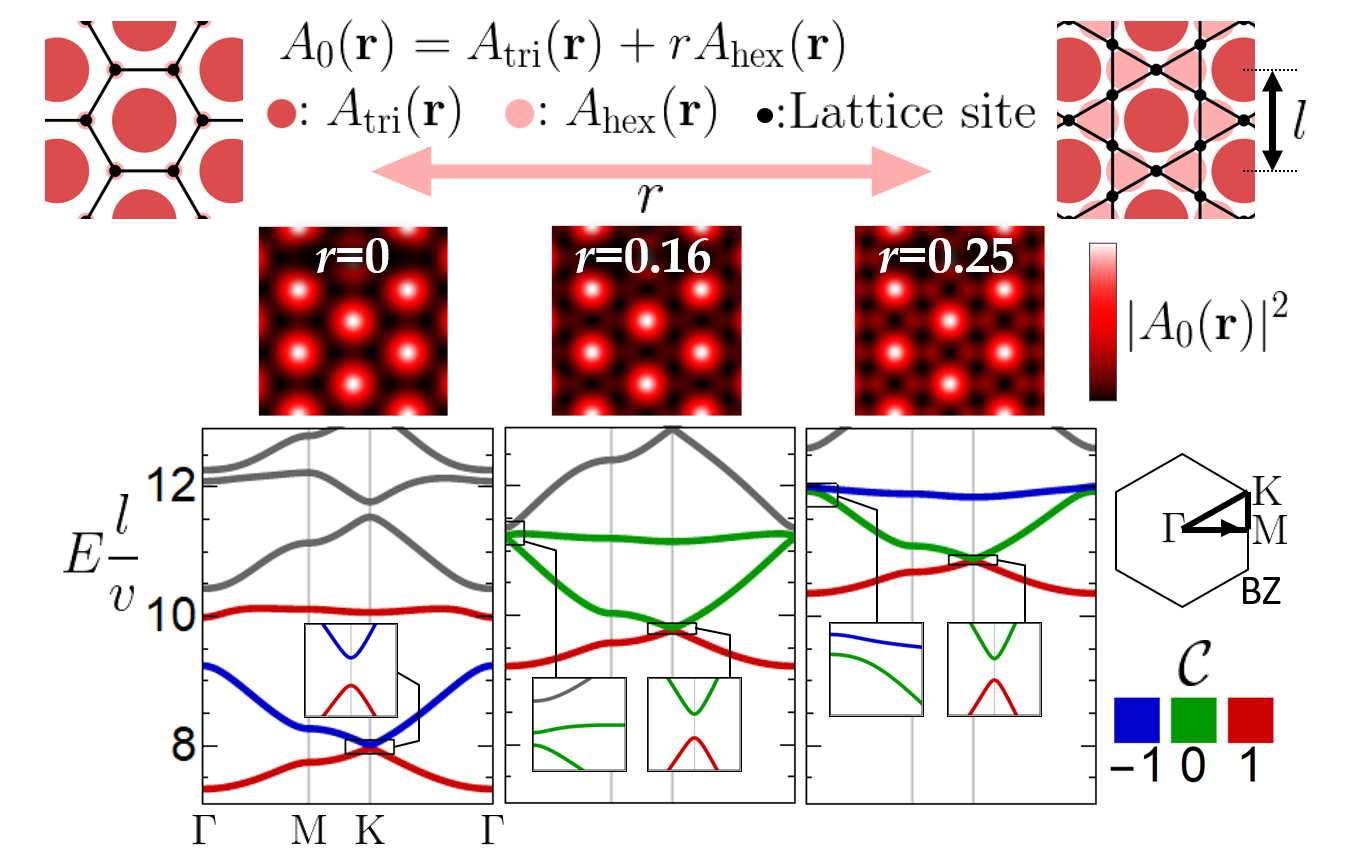}\subfigure{\label{F4a}}\subfigure{\label{F4b}}
\caption{Superposition of the triangular lattice beam, $A_\text{tri}(\mathbf{r})$, and the hexagonal lattice beam $rA_\text{hex}(\mathbf{r})$. As we increase the ratio $r$, we effectively change the electron superlattice from the hexagonal lattice to the kagome lattice. Energy spectra for $\chi=5.4$ at selected values of $r$ are shown where the colors of low-lying bands represent the Chern numbers. By zooming in the spectrum, we can check the gaps in the two-band model and the three-band models in the lowest part of the spectrum.}\label{F4}
\end{figure}

To engineer favorable features such as flatter bands, we can create an even more complicated superlattice by superposing different kinds of lattices. For instance, we consider the superposition of the triangular lattice beam $A_\text{tri}(\mathbf{r})$ and the hexagonal lattice beam $rA_\text{hex}(\mathbf{r})$, where $r$ is the amplitude ratio of the two lattices (\Cref{F4}). When the contribution from the hexagonal lattice beam is negligible, the localized electrons form a hexagonal superlattice and the lowest part of the positive-energy spectrum can be explained by a two-band model. As $r$ increases, electrons are confined to a kagome superlattice \cite{[{In cold-atom systems, optical kagome lattice has been implemented for linearly-polarized lasers as in }]PhysRevLett.108.045305} and the lowest part of the positive-energy spectrum can be explained by a three-band model including a flat band. Note that slight gaps are observed in both the two-band model for the hexagonal superlattice and the three-band model for the kagome superlattice. The gap in the two-band model can be explained with the Haldane model with complex phases in the next-nearest-neighbor tunneling, as shown in \Cref{F3a}. The gap in the kagome lattice comes from the complex phase in the nearest-neighbor tunneling \cite{PhysRevA.82.043811,PhysRevB.93.144307}. At $r=0$, we can see that the third band is nearly flat, while it is gapped well from the other bands. This flat band can be potentially used to stabilize strongly correlated phases.

\section{Experimental feasibility}
For numerical calculation, we have set $\mathcal{A}_0=0.006(ea)^{-1}$, $\omega=0.06t$, and $w/l=0.3$ for \Cref{F2} and \Cref{F3}. With the typical values of $t=3\text{ eV}$ and $a=0.142\text{ nm}$ for the monolayer graphene, these parameters of the laser field correspond to the field amplitude $7.6\times 10^{6}\text{ V/m}$, the beam frequency $43.5\text{ THz}$, and beam spot size $0.1\mu\text{m}$ (FWHM). This is similar to the beam frequency in a recent experiment \cite{mciver2019light} while the peak intensity is about $4\%$ of the beam used in the same experiment. With these parameters, the typical size of the gap ($\Delta_b$ in \Cref{F2}) is 4\text{ meV}. \Cref{F4} uses $\mathcal{A}_0=0.0015(ea)^{-1}$ and $\omega=0.06t$ while $w/l=0.3$ and $w/l=0.15$ for $A_\text{tri}(\mathbf{r})$ and $A_\text{hex}(\mathbf{r})$, respectively. Finally, we remark that due to the injection of photons into the system, heating effects could eventually destroy the nontrivial topological behavior that is initially formed. Therefore, we only consider the prethermal regime where electron-electron and electron-phonon scatterings can be ignored \cite{HDehghani2015Out}. In the past few years, the existence of this transient regime has been convincingly demonstrated  in several pump-probe experiments \cite{wang2013observation, mahmood2016selective, mciver2019light}.

\section{Outlook}
 By considering Coulomb interaction in our nearly flat and topologically nontrivial bands, one could potentially induce strongly correlated phases such as fractional Chern insulators \cite{PhysRevLett.103.196803,PhysRevB.83.195139,maciejko2015fractionalized,spanton2018observation}, superconductors \cite{cao2018unconventional,yankowitz2019tuning,PhysRevX.8.031089,PhysRevLett.121.257001,PhysRevLett.122.257002,PhysRevB.94.214501,*flatsc,martin2019moire}, or magnetic phases \cite{PhysRevLett.119.107201,PhysRevB.98.075109,sharpe2019emergent,PhysRevLett.120.266402}. Moreover, by irradiating with frequencies comparable to the bare tunneling strength, instead of the high-frequency regime considered here, higher-order terms become relevant \cite{PhysRevB.93.144307}, and therefore, one can induce a wider class of structures. While we focus on the Dirac semimetal system in this paper, our scheme can also be applied to other 2D materials such as semiconductors \cite{panna2019ultrafast}. Our approach can be combined with other methods, such as surface acoustic waves in a solid-state platform \cite{PhysRevX.7.041019}, for trapping, cooling, and controlling charged particles, and for simulation of quantum many-body systems. Finally, these ideas could be used to engineer a new class of dielectric materials for potential applications in optical devices \cite{Min:08}.

\section*{Acknowledgments}
I.M. was supported by the Materials Sciences and Engineering Division, Basic Energy Sciences, Office of Science, U.S. Department of Energy. H.A. acknowledges support from JSPS KAKENHI Grant No. 17H06138, and CREST (Core Research for Evolutionary Science and Technology) "Topology" project from JST. H.K., H.D., and M.H. were supported by Grants No. AFOSR FA9550-16-1-0323 and No. FA9550-19-1-0399, Grant No. ARO W911NF2010232, and the NSF Physics Frontier Center at the Joint Quantum Institute. I.M., H.D., and M.H. are thankful for the hospitality of the Kavli Institute for Theoretical Physics, supported by Grant No. NSF PHY-1748958. The authors thank Zhi-Cheng Yang for the insightful discussion.

\appendix
\section{Floquet Effective Hamiltonian in High Frequency Regime}
\setcounter{equation}{0}
\renewcommand{\theequation}{A\arabic{equation}}
Let us consider the Hamiltonian given by \eqnref{H_coupling} with $\mathbf{A}(\mathbf{r},t)=A_0(\mathbf{r})e^{i\omega t}(\mathbf{\hat{x}}+i\mathbf{\hat{y}})+\text{c.c}$. Then we can write the time-dependent Hamiltonian as
\begin{eqnarray}
H(t) &=& v(\tau_z p_x\sigma_x + p_y\sigma_y) +2ev\tau_z A_0(\mathbf{r}) \exp(i\tau_z\omega t) \sigma_+ \nonumber\\
&& +2ev\tau_z A_0(\mathbf{r}) \exp(-i\tau_z\omega t) \sigma_-,
\end{eqnarray}
where $\sigma_\pm = (\sigma_x \pm i\sigma_y)/2$. For this Hamiltonian, the nonzero temporal Fourier components $H_q = (\omega/2\pi)\int_0^{2\pi/\omega} H(t) e^{-iq\omega\tau} d\tau$ are $H_0 = v(\tau_z p_x\sigma_x+p_y\sigma_y)$ and $H_{\pm \tau_z}=2ev\tau_z A_0(\mathbf{r})\sigma_\pm$. Then the effective Hamiltonian in the high frequency regime is \cite{PhysRevB.79.081406,PhysRevLett.110.200403,PhysRevX.4.031027,HDehghani2014Dissipative,HDehghani2015Out,eckardt2015high,PhysRevB.93.144307}
\begin{eqnarray}\label{HFE_ham}
H_\text{eff} &=& H_0 + \sum_{q>0}\frac{[H_q,H_{-q}]}{q\omega} + O(\omega^{-2}) \nonumber\\
&=& H_0 + \frac{H_1 H_{-1} - H_{-1} H_{1}}{\omega} + O(\omega^{-2}) \nonumber\\
&=& v(\tau_z p_x\sigma_x + p_y\sigma_y) + \tau_z\frac{4e^2 v^2}{\omega}\left|A_0(\mathbf{r})\right|^2 \sigma_z + O(\omega^{-2}).\nonumber\\
\end{eqnarray}

The description in terms of \eqnref{HFE_ham} is valid as long as $H_q\ll\omega$ for every $q$. The condition $\omega \ll ev\mathcal{A}_0$ ensures that $H_{q=\pm 1}\ll\omega$. For $H_0 \ll \omega$, we require $v/l\ll\omega$ and the parameters we use in our paper satisfy this condition. Yet, one may wonder if the band structure is affected by the hybridization of different Floquet sidebands \cite{PhysRevB.90.115423,PhysRevB.93.144307} since the driving frequency that we consider in this paper ($\omega=0.06t$) is much smaller than the original bandwidth of the graphene which is of the order of $t$. To see how much our band structure is affected by the Floquet sidebands' hybridization, we calculate the band structure presented in the left figure of \Cref{F2a} by diagonalizing the Floquet Hamiltonian, $H(t)-i\partial_t$. For the spatially periodic Hamiltonian $H(\mathbf{r},t)=H(\mathbf{r}+\mathbf{L}_1,t)=H(\mathbf{r}+\mathbf{L}_2,t)$, we find the quasienergies $\epsilon_{s,\mathbf{k}}$ and the corresponding quasimode wavefunctions $\Psi_{s,\mathbf{k}}(\mathbf{r},t)=\exp(-i\epsilon_s t)\Phi_{s,\mathbf{k}}(\mathbf{r},t)$ through
\begin{eqnarray}\label{FloquetHam}
&& e^{-i\mathbf{k}\cdot\mathbf{r}}\left[H(t)-i\partial_t\right] e^{i\mathbf{k}\cdot\mathbf{r}} \Phi_{s,\mathbf{k}}(\mathbf{r},t)
= \epsilon_{s,\mathbf{k}} \Phi_{s,\mathbf{k}}(\mathbf{r},t),\nonumber\\
&& \Phi_{s,\mathbf{k}}(\mathbf{r},t)
= \sum_{n,m_1,m_2} C_{s,m_1 m_2 \mathbf{k}}^{(n)}
e^{i\lbrace(m_1\mathbf{G}_1 + m_2\mathbf{G}_2 +\mathbf{k})\cdot\mathbf{r}-n\omega t\rbrace},\nonumber\\
&& \sum_{n,m_1,m_2} \left| C_{s,m_1 m_2 \mathbf{k}}^{(n)} \right|^2 = 1,
\end{eqnarray}
where $\mathbf{G}_{i=1,2}$ are the reciprocal superlattice vectors satisfying $\mathbf{G}_i\cdot\mathbf{L}_j = 2\pi\delta_{ij}$. Here, the quasienergies are restricted to the zeroth Floquet sideband, $\epsilon_{s,\mathbf{k}}\in[-\omega/2,\omega/2]$. For the time-independent Hamiltonian $H(t)=H_0$, \eqnref{FloquetHam} becomes an eigenvalue equation for $H_0$ by fixing the Floquet sideband index $n$, and $n=0$ corresponds to the eigenstates with energy in $[-\omega/2,\omega/2]$. Therefore, if we consider the case that oscillating terms are slowly turned on, the relevant quasimodes should have high overlaps with the zeroth Floquet sideband, which is quantified by
\begin{eqnarray}
p_{s,\mathbf{k}}^{(0)} = \sum_{m_1,m_2} \left| C_{s,m_1 m_2 \mathbf{k}}^{(0)} \right|^2.
\end{eqnarray}

\begin{figure}[h]
\centering
\includegraphics[width=\linewidth]{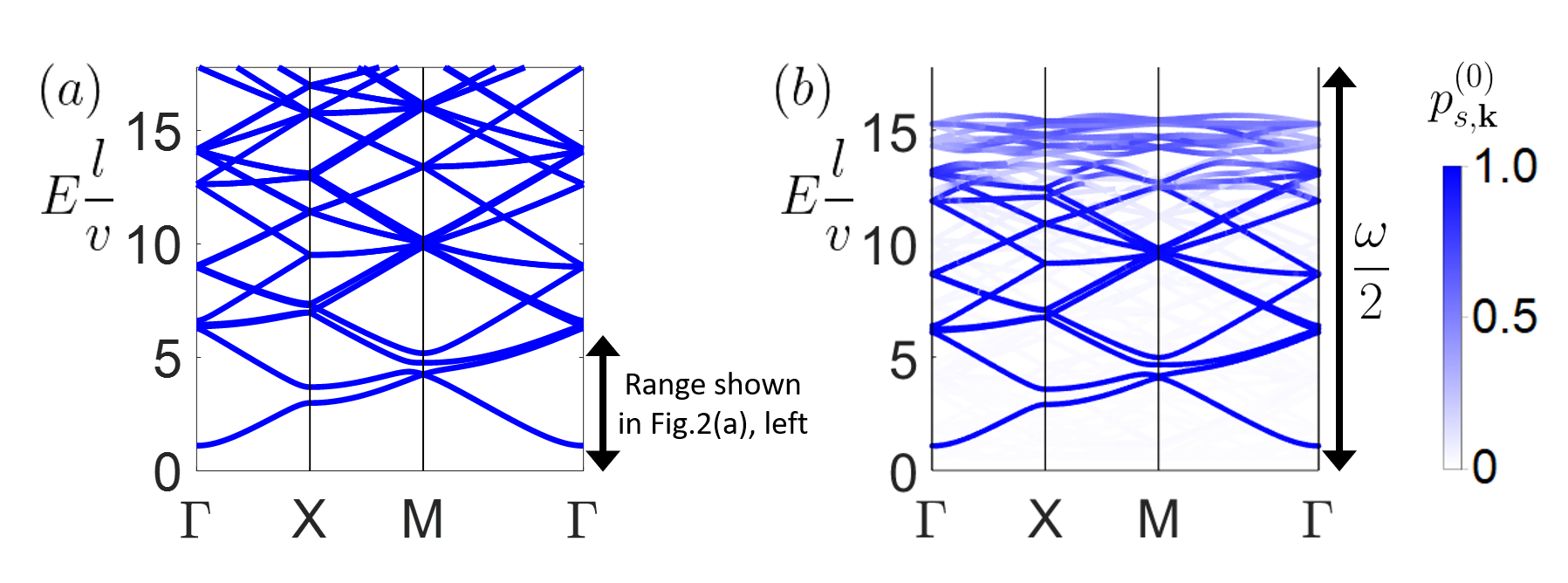}
\caption{Calculated band structure of the square superlattice with $e\mathcal{A}_0 a=0.006$, $\omega=0.06t$, $w/l=0.3$, and $\chi=0.8$. (a) Calculation with \eqnref{HFE_ham}. (b) Calculation with \eqnref{FloquetHam}. The color of the plot represents the overlap with the zeroth Floquet sideband, $p_{s,\mathbf{k}}^{(0)}$.}\label{SF_floquet}
\end{figure}

For the comparison of the two descriptions given by \eqnref{HFE_ham} and \eqnref{FloquetHam}, we calculate the band structure plotted in the left of \Cref{F2a} with these two descriptions, respectively (see \Cref{SF_floquet}). For the band structure calculated with \eqnref{FloquetHam}, we represented the overlap with the zeroth Floquet sideband, $p_{s,\mathbf{k}}^{(0)}$, for each state. In the energy much lower than $\omega/2$, the spectrum calculated with the high-frequency expansion and the quasienergies of the Floquet eigenstates with high overlaps with the zeroth Floquet sideband agree with each other. As the energy approaches $\omega/2$, the Floquet sideband hybridization due to the resonant process affects the band structure. Therefore, one can use the high-frequency expansion description in \eqnref{HFE_ham} for energies far smaller than the driving frequency.

\section{Band Gap Scaling in Superlattice Size}
\setcounter{equation}{0}
\renewcommand{\theequation}{B\arabic{equation}}
We consider the eigenvalue problem of the effective Hamiltonian in \eqnref{eqn_Heff},
\begin{eqnarray}
&&\left( \begin{array}{cc}
\alpha(\mathbf{r}) & -iv(\partial_x -i\partial_y) \\
-iv(\partial_x +i\partial_y) & -\alpha(\mathbf{r})
\end{array} \right)
\left( \begin{array}{c} u_A \\ u_B \end{array} \right)
= E \left( \begin{array}{c} u_A \\ u_B \end{array} \right)
\nonumber\\
&&\leftrightarrow
\begin{array}{c}
-iv(\partial_x-i\partial_y)u_B = [E-\alpha(\mathbf{r})]u_A \\
-iv(\partial_x+i\partial_y)u_A = [E+\alpha(\mathbf{r})]u_B
\end{array}
\end{eqnarray}
where $\alpha(\mathbf{r})=(4e^2v^2/\omega)|A_0(\mathbf{r})|^2$. This can lead to
\begin{eqnarray}
0&=&\nabla^2 u_B +\frac{E^2-\alpha(\mathbf{r})^2}{v^2}u_B \nonumber\\
&&\quad +\frac{[(\partial_x+i\partial_y)\alpha(\mathbf{r})][(\partial_x-i\partial_y)u_B]}{E-\alpha(\mathbf{r})}.
\end{eqnarray}
In the vicinity of minima of $\alpha(\mathbf{r})$, we can approximate this function as a harmonic potential with rotational symmetry. This is valid for the square lattice of the Gaussian beam with the fixed ratio $c=w/l$,
\begin{eqnarray}
\alpha(\mathbf{r}) = \frac{4e^2 v^2 \mathcal{A}_0^2}{\omega} \left[\sum_{n_1,n_2} 
e^{-\lbrace(x-n_1l)^2+(y-n_2l)^2\rbrace/(2c^2l^2)}\right]^2. \qquad\quad
\end{eqnarray}
For this case, one can show that
\begin{eqnarray}
&& \left. \frac{\partial\alpha}{\partial x}\right|_{\mathbf{r}=(l/2,l/2)}
= \left. \frac{\partial\alpha}{\partial y}\right|_{(l/2,l/2)}
= \left. \frac{\partial^2\alpha}{\partial x \partial y}\right|_{(l/2,l/2)}
= 0,
\nonumber\\
&& \alpha_0\equiv \left.\alpha\right|_{(l/2,l/2)}>0,
\alpha_1\equiv \left. \frac{l^2}{2} \partial_x^2\alpha\right|_{(l/2,l/2)} = \left. \frac{l^2}{2} \partial_y^2\alpha\right|_{(l/2,l/2)}>0,\nonumber\\
&& \frac{\partial \alpha_0}{\partial l}=\frac{\partial \alpha_1}{\partial l}=0.
\end{eqnarray}
Then we can write $\alpha(\mathbf{r})=\alpha_0 + \alpha_1 (r/l)^2$, where $r$ is the distance from the minima of $\alpha(\mathbf{r})$. Now we can use polar coordinates $(r,\phi)$, with $\partial_x \pm i\partial_y = a^{-1} e^{\pm i\phi}(\partial_r \pm i r^{-1} \partial_\phi)$. Due to the rotational symmetry, we can impose $u_B(\mathbf{r}) =\beta(r) e^{im\phi}$. Then,
\begin{eqnarray}\label{eqn_rot}
&&\frac{1}{r}\partial_r \left( r\partial_r \beta \right) - \frac{m^2}{r^2} \beta
+\frac{1}{v^2}\left[E^2 - \alpha_0^2 - 2\alpha_0 \alpha_1\left(\frac{r}{l}\right)^2 - \alpha_1^2\left(\frac{r}{l}\right)^4 \right]\beta \nonumber\\
&&\quad +\frac{2\alpha_1 (r\partial_r +m)\beta}{(E-\alpha_0)l^2 -\alpha_1 r^2} = 0.
\end{eqnarray}
Note that $l\to\infty$ limit corresponds to $\nabla^2 u_A + v^{-2} (E^2-\alpha_0^2)=0$. The positive spectrum in this limit is $[\alpha_0,\infty)$ with no gap in between. To study the behavior of the positive spectrum for large $l$, we may define $\delta E = E-\alpha_0$. For the low-lying spectrum, we can only consider the limit where $\delta E \ll \alpha_0$. Then we can simplify \eqnref{eqn_rot} into
\begin{eqnarray}\label{eqn_reduced}
&& \frac{1}{r}\partial_r \left( r\partial_r \beta \right) - \frac{m^2}{r^2} \beta
+\frac{1}{v^2}\left[ 2\alpha_0\delta E - 2\alpha_0 \alpha_1\left(\frac{r}{l}\right)^2 \right]\beta \nonumber\\
&& \quad +\frac{2(r\partial_r +m)\beta}{l^2 \delta E/\alpha_1} = 0
\end{eqnarray}
up to the correction terms of the order of $O(\delta E^2), O(\eta^{-4})$. Now the rescaling $r=(vl)^{1/2}(\alpha_0 \alpha_1)^{-1/4} \xi$ and $\delta E=(\alpha_1/\alpha_0)^{1/2}(v/l)(\delta\epsilon)$ gives
\begin{eqnarray}
&&\frac{1}{\xi}\partial_\xi \left( \xi\partial_\xi \beta \right) - \frac{m^2}{\xi^2} \beta
+2\left( \delta\epsilon - \xi^2 \right)\beta \nonumber\\
&&\quad +\frac{2(\xi\partial_\xi +m)\beta}{\delta\epsilon} = 0
\end{eqnarray}
and this equation is independent of $l$. Then the spectrum of $\delta\epsilon$ is independent of $l$, so that $\delta E$ should scale as $l^{-1}$. This means that $\Delta_b-2\alpha_0$ and $\Delta_t$ should be proportional to $l^{-1}$. This explains the inverse proportionality of band gaps in $\chi=(v e^2 \mathcal{A}_0^2 /
\omega)l$ shown in \Cref{F2b}.

\section{Tight-binding model for hexagonal lattice under a uniform strain}
\setcounter{equation}{0}
\renewcommand{\theequation}{C\arabic{equation}}
\begin{figure}[h]
\centering
\includegraphics[width=0.7\linewidth]{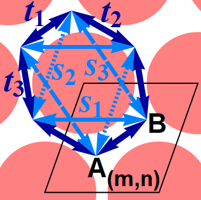}
\caption{Tight-binding model for the hexagonal lattice under a uniform strain, in the vicinity of angle $\theta=\tan^{-1}0.5$ in the sheared lattice.}\label{SF_haldane}
\end{figure}

Let us consider the effective lattice model for the sheared lattice in the vicinity of angle $\theta=\tan^{-1}0.5$. By considering the terms up to the next-nearest neighbors, we can build a tight-binding model similar to the Haldane model,
\begin{eqnarray}\label{H_SH}
H_\text{SH} && = \sum_{m,n}
\left( -t_1 c^{(B)\dag}_{m,n} -t_2 c^{(B)\dag}_{m-1,n} -t_3 c^{(B)\dag}_{m,n-1} \right)c^{(A)}_{m,n} \\
&& +\left( s_1 c^{(A)\dag}_{m+1,n} + s_2 c^{(A)\dag}_{m,n-1} + s_3 c^{(A)\dag}_{m-1,n+1} \right)c^{(A)}_{m,n}  \nonumber\\
&& +\left( s_1 c^{(B)\dag}_{m-1,n} + s_2 c^{(B)\dag}_{m,n+1} + s_3 c^{(B)\dag}_{m+1,n-1} \right)c^{(B)}_{m,n} + \text{H.c.}. \nonumber
\end{eqnarray}
Here, $c^{(A/B)\dag}_{m,n}$ creates an electron in the sublattice $A$ or $B$ at the unit cell $(m,n)$ and $t_{i=1,2,3}$ $(s_{i=1,2,3})$ is the nearest- (next-nearest)-neighbor tunneling amplitude, as shown in \Cref{SF_haldane}. This model can be thought of as a hexagonal lattice under a uniform strain. By considering the inversion symmetry of the corresponding pairs of lattice sites, we can find that $\text{Im}(t_1)=\text{Im}(t_2)=\text{Im}(t_3)=0$. Now we can write the Bloch Hamiltonian of this tight-binding model as
\begin{eqnarray}\label{Bloch}
\mathcal{H}(\mathbf{k}) &=& V(\mathbf{k}) +h_x(\mathbf{k})\sigma_x +h_y(\mathbf{k})\sigma_y +h_z(\mathbf{k})\sigma_z, \nonumber\\
V(\mathbf{k}) &=& 2\text{Re}\left[s_1\cos(\mathbf{k}\cdot(\mathbf{L}_2-\mathbf{L}_1)) +s_2\cos(\mathbf{k}\cdot\mathbf{L}_2)\right.\nonumber\\
&&\left.\qquad +s_3\cos(\mathbf{k}\cdot\mathbf{L}_1)\right],\nonumber\\
h_x(\mathbf{k}) &=& -t_1\cos\left(\mathbf{k}\cdot\frac{\mathbf{L}_1+\mathbf{L}_2}{3} \right) -t_2\cos\left(\mathbf{k}\cdot\frac{\mathbf{L}_2-2\mathbf{L}_1}{3} \right)\nonumber\\
&&\qquad -t_3\cos\left(\mathbf{k}\cdot\frac{\mathbf{L}_1-2\mathbf{L}_2}{3} \right), \nonumber\\
h_y(\mathbf{k}) &=& -t_1\sin\left(\mathbf{k}\cdot\frac{\mathbf{L}_1+\mathbf{L}_2}{3} \right) -t_2\sin\left(\mathbf{k}\cdot\frac{\mathbf{L}_2-2\mathbf{L}_1}{3} \right)\nonumber\\
&&\qquad -t_3\sin\left(\mathbf{k}\cdot\frac{\mathbf{L}_1-2\mathbf{L}_2}{3} \right), \nonumber\\
h_z(\mathbf{k}) &=& 2\text{Im}\left[s_1\sin(\mathbf{k}\cdot(\mathbf{L}_2-\mathbf{L}_1)) +s_2\sin(\mathbf{k}\cdot\mathbf{L}_2)\right.\nonumber\\
&&\left.\qquad +s_3\sin(\mathbf{k}\cdot\mathbf{L}_1)\right],
\end{eqnarray}
where the $\sigma_z=\pm 1$ corresponds to the sublattice $A$ or $B$. In the absence of the next-nearest-neighbor tunnelings, $V=h_z=0$ and the location of the Dirac points is determined by $h_x(\mathbf{k})=h_y(\mathbf{k})=0$. If this equation has two solutions, we denote those solutions as $\mathbf{k}=\pm\mathbf{K}_D$. In \eqnref{Bloch}, we can see that the second-neighbor tunnelings solely determine the $\sigma_z$ component of the Bloch Hamiltonian and do not affect the $\sigma_x$ and $\sigma_y$ components. By turning on the second-neighbor tunnelings, we effectively turn on the mass term around each of the Dirac points $\pm\mathbf{K}_D$. Since $h_z(-\mathbf{K}_D)=-h_z(\mathbf{K}_D)$, the sign of the effective mass term is opposite at the two different Dirac points. Then, each Dirac point equally contributes $1/2$ to the Chern number, just as in the Haldane model. Therefore, if the two Dirac points exist in the absence of the next-nearest-neighbor tunneling, the lowest positive band has nonzero Chern number when the next-nearest-neighbor tunneling is turned on. Regarding this condition, the equation $h_x(\mathbf{k})=h_y(\mathbf{k})=0$ has two solutions as long as $|t_i-t_j|<|t_k|$ for every $i\ne j \ne k \ne i$. Since the tunneling strength decreases exponentially in the intersite distance, $t_1$, $t_2$, and $t_3$ become very different as the superlattice size gets larger. Then there is no Dirac point after some value of $\chi$, as shown in the square-lattice case. Yet, at angle $\theta=\tan^{-1} 0.5$, $t_1=t_2$, so that $|t_i-t_j|<|t_k|$ is satisfied as long as $t_3\ne 0$, and therefore $\mathcal{C}_1$ can remain nonzero at this angle.

\section{Gauge-Independent Calculation of Orbital Magnetization}
\setcounter{equation}{0}
\renewcommand{\theequation}{D\arabic{equation}}
We want to numerically calculate the orbital magnetization of the $m$th band expressed in \eqnref{eqn_Morb},
\begin{eqnarray}\label{seqn_Morb}
M_\text{orb}&=& \text{Im}\int \frac{d^2\mathbf{k}}{(2\pi)^2}
e\frac{\partial\bra{u_{m,\mathbf{k}}}}{\partial_{k_x}}
\left( H_{\mathbf{k}} + E_{m,\mathbf{k}} \right)
\frac{\partial\ket{u_{m,\mathbf{k}}}}{\partial_{k_y}} \nonumber\\
&=& \text{Im}\int \frac{d^2\mathbf{k}}{(2\pi)^2}
e\frac{\partial\bra{u_{m,\mathbf{k}}}}{\partial_{k_x}}
H_{\mathbf{k}} \frac{\partial\ket{u_{m,\mathbf{k}}}}{\partial_{k_y}} \nonumber\\
&&\quad + \frac{1}{2}\int \frac{d^2\mathbf{k}}{(2\pi)^2}
e E_{m,\mathbf{k}} \mathcal{A}_{m,\mathbf{k}},
\end{eqnarray}
where $\mathcal{A}_{m,\mathbf{k}}=2\text{Im}\left(\partial_{k_x} \bra{ u_{m,\mathbf{k}_j}} \right)\partial_{k_y}\ket{ u_{m,\mathbf{k}_j}}$ is the Berry curvature. To numerically calculate it, we first need to discretize the Brillouin zone and calculate the Bloch state $\ket{\psi_{m,\mathbf{k}_j}}$. Although the orbital magnetization is gauge independent, we need local gauge fixing to make $\ket{\psi_{m,\mathbf{k}}}$ differentiable. While this local gauge fixing works well for smooth $\mathcal{A}_{m,\mathbf{k}}$, it can work badly for the system in the vicinity of the topological phase transition. To avoid this subtlety, let us find a way to calculate this quantity in a gauge-independent way. For Berry curvature $\mathcal{A}_{m,\mathbf{k}}$, a method for gauge-independent calculation is known \cite{fukui2005chern}. Similar to this method, we can calculate the first integral of \eqnref{seqn_Morb}. For this, let us consider a square patch whose four corners are $\mathbf{q}_{l,j}\equiv\mathbf{k}_l +(\delta k /2)(s_{j}\mathbf{\hat{x}}+w_{j}\mathbf{\hat{y}})$, where $(s_1,w_1)=(-1,-1)$, $(s_2,w_2)=(1,-1)$, $(s_3,w_3)=(1,1)$, and $(s_4,w_4)=(-1,1)$. Now
\begin{eqnarray}
&& \ket{u_{m,\mathbf{q}_{l,j}}} = \ket{u_{m,\mathbf{k}_l}}
+\frac{\delta k}{2}\left( s_j\partial_{k_x}\ket{ u_{m,\mathbf{k}_l}} + w_j\partial_{k_y}\ket{ u_{m,\mathbf{k}_l}} \right) \nonumber\\
&&\quad
+\frac{\delta k^2}{8}\left( \partial_{k_x}^2\ket{ u_{m,\mathbf{k}_l}} + 2 s_j w_j\partial_{k_x}\partial_{k_y}\ket{ u_{m,\mathbf{k}_l}} + \partial_{k_y}^2\ket{ u_{m,\mathbf{k}_l}} \right) \nonumber\\
&&\quad +O(\delta k^3), \nonumber\\
&& E_{m,\mathbf{k}_l}^{-4} \prod_{j=1}^4 \braket{ u_{m,\mathbf{q}_{l,j}} |H_{\mathbf{k}_l}| u_{m,\mathbf{q}_{l,(j\text{mod}4)+1}} } \nonumber\\
&&= 1 + \delta k \text{Re} \sum_{j=1}^4\left( s_j \braket{u_{m,\mathbf{k}_l}|\partial_{k_x}|u_{m,\mathbf{k}_l}} + w_j\braket{u_{m,\mathbf{k}_l}|\partial_{k_y}|u_{m,\mathbf{k}_l}} \right) \nonumber\\
&&\quad +\delta k^2 \text{Re} \braket{u_{m,\mathbf{k}_l}|\nabla_\mathbf{k}^2|u_{m,\mathbf{k}_l}} \nonumber\\
&&\quad +\delta k^2 \text{Re}\sum_{j} \frac{s_j w_j}{2} \braket{u_{m,\mathbf{k}_l}|\partial_{k_x}\partial_{k_y}|u_{m,\mathbf{k}_l}} \nonumber\\
&&\quad +\frac{\delta k^2}{4 E_{m,\mathbf{k}_l}} \sum_{j=1}^4\left[
s_j s_{(j\text{mod}4)+1} \left( \partial_{k_x} \bra{u_{m,\mathbf{k}_l}} \right) H_{\mathbf{k}_l}\partial_{k_x}\ket{ u_{m,\mathbf{k}_l}} \right.\nonumber\\
&&\quad
+ s_j w_{(j\text{mod}4)+1} \left( \partial_{k_x} \bra{u_{m,\mathbf{k}_l}} \right) H_{\mathbf{k}_l}\partial_{k_y}\ket{ u_{m,\mathbf{k}_l}} \nonumber\\
&&\quad
+ w_j s_{(j\text{mod}4)+1} \left( \partial_{k_y} \bra{u_{m,\mathbf{k}_l}} \right) H_{\mathbf{k}_l}\partial_{k_x}\ket{ u_{m,\mathbf{k}_l}} \nonumber\\
&&\quad\left.
+ w_j w_{(j\text{mod}4)+1} \left( \partial_{k_y} \bra{u_{m,\mathbf{k}_l}} \right) H_{\mathbf{k}_l}\partial_{k_y}\ket{ u_{m,\mathbf{k}_l}}
\right] +O(\delta k^3) \nonumber\\
&&= 1 + 2\delta k^2 \text{Re}\braket{u_{m,\mathbf{k}_l}|\nabla_\mathbf{k}^2|u_{m,\mathbf{k}_l}} \nonumber\\
&&\quad +i\frac{2\delta k^2}{E_{m,\mathbf{k}_l}} \text{Im} \left( \partial_{k_x}\bra{ u_{m,\mathbf{k}_l}}\right) H_{\mathbf{k}_l}\partial_{k_y}\ket{ u_{m,\mathbf{k}_l}} +O(\delta k^3),
\end{eqnarray}
and therefore
\begin{eqnarray}
&& \frac{e E_{m,\mathbf{k}_l}}{8\pi^2}\text{Arg} \left( \prod_{j=1}^4 \braket{ u_{m,\mathbf{q}_{l,j}} |H_{\mathbf{k}_l}| u_{m,\mathbf{q}_{l,(j\text{mod}4)+1}} } \right) \nonumber\\
&& = \frac{e}{4\pi^2} \text{Im} \left( \partial_{k_x}\bra{ u_{m,\mathbf{k}_l}}\right) H_{\mathbf{k}_l}\partial_{k_y}\ket{ u_{m,\mathbf{k}_l}} \delta k^2 +O(\delta k^3),\qquad\quad
\end{eqnarray}
and this corresponds to the first integral of \eqnref{seqn_Morb} over the square patch that we considered. One can easily check that this expression is invariant under any gauge transformation, $\ket{u_{m,\mathbf{k}}}\to \exp[i\lambda(\mathbf{k})] \ket{u_{m,\mathbf{k}}},\ \forall\lambda(\mathbf{k})$, and does not require any local gauge fixing.
\\\\\\\\\\\\\\\\\\\\\\\\\\\\\\\\\\\\\\\\\\\\\\\\\\\\\\\
%


\begin{thebibliography}{70}%
\makeatletter
\providecommand \@ifxundefined [1]{%
 \@ifx{#1\undefined}
}%
\providecommand \@ifnum [1]{%
 \ifnum #1\expandafter \@firstoftwo
 \else \expandafter \@secondoftwo
 \fi
}%
\providecommand \@ifx [1]{%
 \ifx #1\expandafter \@firstoftwo
 \else \expandafter \@secondoftwo
 \fi
}%
\providecommand \natexlab [1]{#1}%
\providecommand \enquote  [1]{``#1''}%
\providecommand \bibnamefont  [1]{#1}%
\providecommand \bibfnamefont [1]{#1}%
\providecommand \citenamefont [1]{#1}%
\providecommand \href@noop [0]{\@secondoftwo}%
\providecommand \href [0]{\begingroup \@sanitize@url \@href}%
\providecommand \@href[1]{\@@startlink{#1}\@@href}%
\providecommand \@@href[1]{\endgroup#1\@@endlink}%
\providecommand \@sanitize@url [0]{\catcode `\\12\catcode `\$12\catcode
  `\&12\catcode `\#12\catcode `\^12\catcode `\_12\catcode `\%12\relax}%
\providecommand \@@startlink[1]{}%
\providecommand \@@endlink[0]{}%
\providecommand \url  [0]{\begingroup\@sanitize@url \@url }%
\providecommand \@url [1]{\endgroup\@href {#1}{\urlprefix }}%
\providecommand \urlprefix  [0]{URL }%
\providecommand \Eprint [0]{\href }%
\providecommand \doibase [0]{http://dx.doi.org/}%
\providecommand \selectlanguage [0]{\@gobble}%
\providecommand \bibinfo  [0]{\@secondoftwo}%
\providecommand \bibfield  [0]{\@secondoftwo}%
\providecommand \translation [1]{[#1]}%
\providecommand \BibitemOpen [0]{}%
\providecommand \bibitemStop [0]{}%
\providecommand \bibitemNoStop [0]{.\EOS\space}%
\providecommand \EOS [0]{\spacefactor3000\relax}%
\providecommand \BibitemShut  [1]{\csname bibitem#1\endcsname}%
\let\auto@bib@innerbib\@empty
\bibitem [{\citenamefont {Shima}\ and\ \citenamefont
  {Aoki}(1993)}]{PhysRevLett.71.4389}%
  \BibitemOpen
  \bibfield  {author} {\bibinfo {author} {\bibfnamefont {N.}~\bibnamefont
  {Shima}}\ and\ \bibinfo {author} {\bibfnamefont {H.}~\bibnamefont {Aoki}},\
  }\href {\doibase 10.1103/PhysRevLett.71.4389} {\bibfield  {journal} {\bibinfo
   {journal} {Phys. Rev. Lett.}\ }\textbf {\bibinfo {volume} {71}},\ \bibinfo
  {pages} {4389} (\bibinfo {year} {1993})}\BibitemShut {NoStop}%
\bibitem [{\citenamefont {Bistritzer}\ and\ \citenamefont
  {MacDonald}(2011{\natexlab{a}})}]{PhysRevB.84.035440}%
  \BibitemOpen
  \bibfield  {author} {\bibinfo {author} {\bibfnamefont {R.}~\bibnamefont
  {Bistritzer}}\ and\ \bibinfo {author} {\bibfnamefont {A.~H.}\ \bibnamefont
  {MacDonald}},\ }\href {\doibase 10.1103/PhysRevB.84.035440} {\bibfield
  {journal} {\bibinfo  {journal} {Phys. Rev. B}\ }\textbf {\bibinfo {volume}
  {84}},\ \bibinfo {pages} {035440} (\bibinfo {year}
  {2011}{\natexlab{a}})}\BibitemShut {NoStop}%
\bibitem [{\citenamefont {Spanton}\ \emph {et~al.}(2018)\citenamefont
  {Spanton}, \citenamefont {Zibrov}, \citenamefont {Zhou}, \citenamefont
  {Taniguchi}, \citenamefont {Watanabe}, \citenamefont {Zaletel},\ and\
  \citenamefont {Young}}]{spanton2018observation}%
  \BibitemOpen
  \bibfield  {author} {\bibinfo {author} {\bibfnamefont {E.~M.}\ \bibnamefont
  {Spanton}}, \bibinfo {author} {\bibfnamefont {A.~A.}\ \bibnamefont {Zibrov}},
  \bibinfo {author} {\bibfnamefont {H.}~\bibnamefont {Zhou}}, \bibinfo {author}
  {\bibfnamefont {T.}~\bibnamefont {Taniguchi}}, \bibinfo {author}
  {\bibfnamefont {K.}~\bibnamefont {Watanabe}}, \bibinfo {author}
  {\bibfnamefont {M.~P.}\ \bibnamefont {Zaletel}}, \ and\ \bibinfo {author}
  {\bibfnamefont {A.~F.}\ \bibnamefont {Young}},\ }\href
  {https://science.sciencemag.org/content/360/6384/62} {\bibfield  {journal}
  {\bibinfo  {journal} {Science}\ }\textbf {\bibinfo {volume} {360}},\ \bibinfo
  {pages} {62} (\bibinfo {year} {2018})}\BibitemShut {NoStop}%
\bibitem [{\citenamefont {Bistritzer}\ and\ \citenamefont
  {MacDonald}(2011{\natexlab{b}})}]{bistritzer2011moire}%
  \BibitemOpen
  \bibfield  {author} {\bibinfo {author} {\bibfnamefont {R.}~\bibnamefont
  {Bistritzer}}\ and\ \bibinfo {author} {\bibfnamefont {A.~H.}\ \bibnamefont
  {MacDonald}},\ }\href {https://www.pnas.org/content/108/30/12233.short}
  {\bibfield  {journal} {\bibinfo  {journal} {Proceedings of the National
  Academy of Sciences}\ }\textbf {\bibinfo {volume} {108}},\ \bibinfo {pages}
  {12233} (\bibinfo {year} {2011}{\natexlab{b}})}\BibitemShut {NoStop}%
\bibitem [{\citenamefont {Lopes~dos Santos}\ \emph {et~al.}(2012)\citenamefont
  {Lopes~dos Santos}, \citenamefont {Peres},\ and\ \citenamefont
  {Castro~Neto}}]{PhysRevB.86.155449}%
  \BibitemOpen
  \bibfield  {author} {\bibinfo {author} {\bibfnamefont {J.~M.~B.}\
  \bibnamefont {Lopes~dos Santos}}, \bibinfo {author} {\bibfnamefont
  {N.~M.~R.}\ \bibnamefont {Peres}}, \ and\ \bibinfo {author} {\bibfnamefont
  {A.~H.}\ \bibnamefont {Castro~Neto}},\ }\href {\doibase
  10.1103/PhysRevB.86.155449} {\bibfield  {journal} {\bibinfo  {journal} {Phys.
  Rev. B}\ }\textbf {\bibinfo {volume} {86}},\ \bibinfo {pages} {155449}
  (\bibinfo {year} {2012})}\BibitemShut {NoStop}%
\bibitem [{\citenamefont {Tarnopolsky}\ \emph {et~al.}(2019)\citenamefont
  {Tarnopolsky}, \citenamefont {Kruchkov},\ and\ \citenamefont
  {Vishwanath}}]{PhysRevLett.122.106405}%
  \BibitemOpen
  \bibfield  {author} {\bibinfo {author} {\bibfnamefont {G.}~\bibnamefont
  {Tarnopolsky}}, \bibinfo {author} {\bibfnamefont {A.~J.}\ \bibnamefont
  {Kruchkov}}, \ and\ \bibinfo {author} {\bibfnamefont {A.}~\bibnamefont
  {Vishwanath}},\ }\href {\doibase 10.1103/PhysRevLett.122.106405} {\bibfield
  {journal} {\bibinfo  {journal} {Phys. Rev. Lett.}\ }\textbf {\bibinfo
  {volume} {122}},\ \bibinfo {pages} {106405} (\bibinfo {year}
  {2019})}\BibitemShut {NoStop}%
\bibitem [{\citenamefont {Cao}\ \emph {et~al.}(2018{\natexlab{a}})\citenamefont
  {Cao}, \citenamefont {Fatemi}, \citenamefont {Fang}, \citenamefont
  {Watanabe}, \citenamefont {Taniguchi}, \citenamefont {Kaxiras},\ and\
  \citenamefont {Jarillo-Herrero}}]{cao2018unconventional}%
  \BibitemOpen
  \bibfield  {author} {\bibinfo {author} {\bibfnamefont {Y.}~\bibnamefont
  {Cao}}, \bibinfo {author} {\bibfnamefont {V.}~\bibnamefont {Fatemi}},
  \bibinfo {author} {\bibfnamefont {S.}~\bibnamefont {Fang}}, \bibinfo {author}
  {\bibfnamefont {K.}~\bibnamefont {Watanabe}}, \bibinfo {author}
  {\bibfnamefont {T.}~\bibnamefont {Taniguchi}}, \bibinfo {author}
  {\bibfnamefont {E.}~\bibnamefont {Kaxiras}}, \ and\ \bibinfo {author}
  {\bibfnamefont {P.}~\bibnamefont {Jarillo-Herrero}},\ }\href
  {https://www.nature.com/articles/nature26160} {\bibfield  {journal} {\bibinfo
   {journal} {Nature}\ }\textbf {\bibinfo {volume} {556}},\ \bibinfo {pages}
  {43} (\bibinfo {year} {2018}{\natexlab{a}})}\BibitemShut {NoStop}%
\bibitem [{\citenamefont {Cao}\ \emph {et~al.}(2018{\natexlab{b}})\citenamefont
  {Cao}, \citenamefont {Fatemi}, \citenamefont {Demir}, \citenamefont {Fang},
  \citenamefont {Tomarken}, \citenamefont {Luo}, \citenamefont
  {Sanchez-Yamagishi}, \citenamefont {Watanabe}, \citenamefont {Taniguchi},
  \citenamefont {Kaxiras}, \citenamefont {Ashoori},\ and\ \citenamefont
  {Jarillo-Herrero}}]{cao2018correlated}%
  \BibitemOpen
  \bibfield  {author} {\bibinfo {author} {\bibfnamefont {Y.}~\bibnamefont
  {Cao}}, \bibinfo {author} {\bibfnamefont {V.}~\bibnamefont {Fatemi}},
  \bibinfo {author} {\bibfnamefont {A.}~\bibnamefont {Demir}}, \bibinfo
  {author} {\bibfnamefont {S.}~\bibnamefont {Fang}}, \bibinfo {author}
  {\bibfnamefont {S.~L.}\ \bibnamefont {Tomarken}}, \bibinfo {author}
  {\bibfnamefont {J.~Y.}\ \bibnamefont {Luo}}, \bibinfo {author} {\bibfnamefont
  {J.~D.}\ \bibnamefont {Sanchez-Yamagishi}}, \bibinfo {author} {\bibfnamefont
  {K.}~\bibnamefont {Watanabe}}, \bibinfo {author} {\bibfnamefont
  {T.}~\bibnamefont {Taniguchi}}, \bibinfo {author} {\bibfnamefont
  {E.}~\bibnamefont {Kaxiras}}, \bibinfo {author} {\bibfnamefont {R.~C.}\
  \bibnamefont {Ashoori}}, \ and\ \bibinfo {author} {\bibfnamefont
  {P.}~\bibnamefont {Jarillo-Herrero}},\ }\href
  {https://www.nature.com/articles/nature26154} {\bibfield  {journal} {\bibinfo
   {journal} {Nature}\ }\textbf {\bibinfo {volume} {556}},\ \bibinfo {pages}
  {80} (\bibinfo {year} {2018}{\natexlab{b}})}\BibitemShut {NoStop}%
\bibitem [{\citenamefont {Sharpe}\ \emph {et~al.}(2019)\citenamefont {Sharpe},
  \citenamefont {Fox}, \citenamefont {Barnard}, \citenamefont {Finney},
  \citenamefont {Watanabe}, \citenamefont {Taniguchi}, \citenamefont
  {Kastner},\ and\ \citenamefont {Goldhaber-Gordon}}]{sharpe2019emergent}%
  \BibitemOpen
  \bibfield  {author} {\bibinfo {author} {\bibfnamefont {A.~L.}\ \bibnamefont
  {Sharpe}}, \bibinfo {author} {\bibfnamefont {E.~J.}\ \bibnamefont {Fox}},
  \bibinfo {author} {\bibfnamefont {A.~W.}\ \bibnamefont {Barnard}}, \bibinfo
  {author} {\bibfnamefont {J.}~\bibnamefont {Finney}}, \bibinfo {author}
  {\bibfnamefont {K.}~\bibnamefont {Watanabe}}, \bibinfo {author}
  {\bibfnamefont {T.}~\bibnamefont {Taniguchi}}, \bibinfo {author}
  {\bibfnamefont {M.}~\bibnamefont {Kastner}}, \ and\ \bibinfo {author}
  {\bibfnamefont {D.}~\bibnamefont {Goldhaber-Gordon}},\ }\href
  {https://science.sciencemag.org/content/365/6453/605} {\bibfield  {journal}
  {\bibinfo  {journal} {Science}\ }\textbf {\bibinfo {volume} {365}},\ \bibinfo
  {pages} {605} (\bibinfo {year} {2019})}\BibitemShut {NoStop}%
\bibitem [{\citenamefont {Liu}\ \emph {et~al.}(2019)\citenamefont {Liu},
  \citenamefont {Ma}, \citenamefont {Gao},\ and\ \citenamefont
  {Dai}}]{PhysRevX.9.031021}%
  \BibitemOpen
  \bibfield  {author} {\bibinfo {author} {\bibfnamefont {J.}~\bibnamefont
  {Liu}}, \bibinfo {author} {\bibfnamefont {Z.}~\bibnamefont {Ma}}, \bibinfo
  {author} {\bibfnamefont {J.}~\bibnamefont {Gao}}, \ and\ \bibinfo {author}
  {\bibfnamefont {X.}~\bibnamefont {Dai}},\ }\href {\doibase
  10.1103/PhysRevX.9.031021} {\bibfield  {journal} {\bibinfo  {journal} {Phys.
  Rev. X}\ }\textbf {\bibinfo {volume} {9}},\ \bibinfo {pages} {031021}
  (\bibinfo {year} {2019})}\BibitemShut {NoStop}%
\bibitem [{\citenamefont {Choi}\ \emph {et~al.}(2019)\citenamefont {Choi},
  \citenamefont {Kemmer}, \citenamefont {Peng}, \citenamefont {Thomson},
  \citenamefont {Arora}, \citenamefont {Polski}, \citenamefont {Zhang},
  \citenamefont {Ren}, \citenamefont {Alicea}, \citenamefont {Refael},
  \citenamefont {Oppen}, \citenamefont {Watanabe}, \citenamefont {Taniguchi},\
  and\ \citenamefont {Nadj-Perge}}]{choi2019electronic}%
  \BibitemOpen
  \bibfield  {author} {\bibinfo {author} {\bibfnamefont {Y.}~\bibnamefont
  {Choi}}, \bibinfo {author} {\bibfnamefont {J.}~\bibnamefont {Kemmer}},
  \bibinfo {author} {\bibfnamefont {Y.}~\bibnamefont {Peng}}, \bibinfo {author}
  {\bibfnamefont {A.}~\bibnamefont {Thomson}}, \bibinfo {author} {\bibfnamefont
  {H.}~\bibnamefont {Arora}}, \bibinfo {author} {\bibfnamefont
  {R.}~\bibnamefont {Polski}}, \bibinfo {author} {\bibfnamefont
  {Y.}~\bibnamefont {Zhang}}, \bibinfo {author} {\bibfnamefont
  {H.}~\bibnamefont {Ren}}, \bibinfo {author} {\bibfnamefont {J.}~\bibnamefont
  {Alicea}}, \bibinfo {author} {\bibfnamefont {G.}~\bibnamefont {Refael}},
  \bibinfo {author} {\bibfnamefont {F.~v.}\ \bibnamefont {Oppen}}, \bibinfo
  {author} {\bibfnamefont {K.}~\bibnamefont {Watanabe}}, \bibinfo {author}
  {\bibfnamefont {T.}~\bibnamefont {Taniguchi}}, \ and\ \bibinfo {author}
  {\bibfnamefont {S.}~\bibnamefont {Nadj-Perge}},\ }\href
  {https://doi.org/10.1038/s41567-019-0606-5} {\bibfield  {journal} {\bibinfo
  {journal} {Nat. Phys.}\ }\textbf {\bibinfo {volume} {15}},\ \bibinfo {pages}
  {1174} (\bibinfo {year} {2019})}\BibitemShut {NoStop}%
\bibitem [{\citenamefont {Chen}\ \emph {et~al.}(2019)\citenamefont {Chen},
  \citenamefont {Jiang}, \citenamefont {Wu}, \citenamefont {Lyu}, \citenamefont
  {Li}, \citenamefont {Chittari}, \citenamefont {Watanabe}, \citenamefont
  {Taniguchi}, \citenamefont {Shi}, \citenamefont {Jung}, \citenamefont
  {Zhang},\ and\ \citenamefont {Wang}}]{chen2019evidence}%
  \BibitemOpen
  \bibfield  {author} {\bibinfo {author} {\bibfnamefont {G.}~\bibnamefont
  {Chen}}, \bibinfo {author} {\bibfnamefont {L.}~\bibnamefont {Jiang}},
  \bibinfo {author} {\bibfnamefont {S.}~\bibnamefont {Wu}}, \bibinfo {author}
  {\bibfnamefont {B.}~\bibnamefont {Lyu}}, \bibinfo {author} {\bibfnamefont
  {H.}~\bibnamefont {Li}}, \bibinfo {author} {\bibfnamefont {B.~L.}\
  \bibnamefont {Chittari}}, \bibinfo {author} {\bibfnamefont {K.}~\bibnamefont
  {Watanabe}}, \bibinfo {author} {\bibfnamefont {T.}~\bibnamefont {Taniguchi}},
  \bibinfo {author} {\bibfnamefont {Z.}~\bibnamefont {Shi}}, \bibinfo {author}
  {\bibfnamefont {J.}~\bibnamefont {Jung}}, \bibinfo {author} {\bibfnamefont
  {Y.}~\bibnamefont {Zhang}}, \ and\ \bibinfo {author} {\bibfnamefont
  {F.}~\bibnamefont {Wang}},\ }\href
  {https://www.nature.com/articles/s41567-018-0387-2} {\bibfield  {journal}
  {\bibinfo  {journal} {Nature Physics}\ }\textbf {\bibinfo {volume} {15}},\
  \bibinfo {pages} {237} (\bibinfo {year} {2019})}\BibitemShut {NoStop}%
\bibitem [{\citenamefont {Po}\ \emph {et~al.}(2018)\citenamefont {Po},
  \citenamefont {Zou}, \citenamefont {Vishwanath},\ and\ \citenamefont
  {Senthil}}]{PhysRevX.8.031089}%
  \BibitemOpen
  \bibfield  {author} {\bibinfo {author} {\bibfnamefont {H.~C.}\ \bibnamefont
  {Po}}, \bibinfo {author} {\bibfnamefont {L.}~\bibnamefont {Zou}}, \bibinfo
  {author} {\bibfnamefont {A.}~\bibnamefont {Vishwanath}}, \ and\ \bibinfo
  {author} {\bibfnamefont {T.}~\bibnamefont {Senthil}},\ }\href {\doibase
  10.1103/PhysRevX.8.031089} {\bibfield  {journal} {\bibinfo  {journal} {Phys.
  Rev. X}\ }\textbf {\bibinfo {volume} {8}},\ \bibinfo {pages} {031089}
  (\bibinfo {year} {2018})}\BibitemShut {NoStop}%
\bibitem [{\citenamefont {Gonzalez-Arraga}\ \emph {et~al.}(2017)\citenamefont
  {Gonzalez-Arraga}, \citenamefont {Lado}, \citenamefont {Guinea},\ and\
  \citenamefont {San-Jose}}]{PhysRevLett.119.107201}%
  \BibitemOpen
  \bibfield  {author} {\bibinfo {author} {\bibfnamefont {L.~A.}\ \bibnamefont
  {Gonzalez-Arraga}}, \bibinfo {author} {\bibfnamefont {J.~L.}\ \bibnamefont
  {Lado}}, \bibinfo {author} {\bibfnamefont {F.}~\bibnamefont {Guinea}}, \ and\
  \bibinfo {author} {\bibfnamefont {P.}~\bibnamefont {San-Jose}},\ }\href
  {\doibase 10.1103/PhysRevLett.119.107201} {\bibfield  {journal} {\bibinfo
  {journal} {Phys. Rev. Lett.}\ }\textbf {\bibinfo {volume} {119}},\ \bibinfo
  {pages} {107201} (\bibinfo {year} {2017})}\BibitemShut {NoStop}%
\bibitem [{\citenamefont {Thomson}\ \emph {et~al.}(2018)\citenamefont
  {Thomson}, \citenamefont {Chatterjee}, \citenamefont {Sachdev},\ and\
  \citenamefont {Scheurer}}]{PhysRevB.98.075109}%
  \BibitemOpen
  \bibfield  {author} {\bibinfo {author} {\bibfnamefont {A.}~\bibnamefont
  {Thomson}}, \bibinfo {author} {\bibfnamefont {S.}~\bibnamefont {Chatterjee}},
  \bibinfo {author} {\bibfnamefont {S.}~\bibnamefont {Sachdev}}, \ and\
  \bibinfo {author} {\bibfnamefont {M.~S.}\ \bibnamefont {Scheurer}},\ }\href
  {\doibase 10.1103/PhysRevB.98.075109} {\bibfield  {journal} {\bibinfo
  {journal} {Phys. Rev. B}\ }\textbf {\bibinfo {volume} {98}},\ \bibinfo
  {pages} {075109} (\bibinfo {year} {2018})}\BibitemShut {NoStop}%
\bibitem [{\citenamefont {Sboychakov}\ \emph {et~al.}(2018)\citenamefont
  {Sboychakov}, \citenamefont {Rozhkov}, \citenamefont {Rakhmanov},\ and\
  \citenamefont {Nori}}]{PhysRevLett.120.266402}%
  \BibitemOpen
  \bibfield  {author} {\bibinfo {author} {\bibfnamefont {A.~O.}\ \bibnamefont
  {Sboychakov}}, \bibinfo {author} {\bibfnamefont {A.~V.}\ \bibnamefont
  {Rozhkov}}, \bibinfo {author} {\bibfnamefont {A.~L.}\ \bibnamefont
  {Rakhmanov}}, \ and\ \bibinfo {author} {\bibfnamefont {F.}~\bibnamefont
  {Nori}},\ }\href {\doibase 10.1103/PhysRevLett.120.266402} {\bibfield
  {journal} {\bibinfo  {journal} {Phys. Rev. Lett.}\ }\textbf {\bibinfo
  {volume} {120}},\ \bibinfo {pages} {266402} (\bibinfo {year}
  {2018})}\BibitemShut {NoStop}%
\bibitem [{\citenamefont {Yankowitz}\ \emph {et~al.}(2019)\citenamefont
  {Yankowitz}, \citenamefont {Chen}, \citenamefont {Polshyn}, \citenamefont
  {Zhang}, \citenamefont {Watanabe}, \citenamefont {Taniguchi}, \citenamefont
  {Graf}, \citenamefont {Young},\ and\ \citenamefont
  {Dean}}]{yankowitz2019tuning}%
  \BibitemOpen
  \bibfield  {author} {\bibinfo {author} {\bibfnamefont {M.}~\bibnamefont
  {Yankowitz}}, \bibinfo {author} {\bibfnamefont {S.}~\bibnamefont {Chen}},
  \bibinfo {author} {\bibfnamefont {H.}~\bibnamefont {Polshyn}}, \bibinfo
  {author} {\bibfnamefont {Y.}~\bibnamefont {Zhang}}, \bibinfo {author}
  {\bibfnamefont {K.}~\bibnamefont {Watanabe}}, \bibinfo {author}
  {\bibfnamefont {T.}~\bibnamefont {Taniguchi}}, \bibinfo {author}
  {\bibfnamefont {D.}~\bibnamefont {Graf}}, \bibinfo {author} {\bibfnamefont
  {A.~F.}\ \bibnamefont {Young}}, \ and\ \bibinfo {author} {\bibfnamefont
  {C.~R.}\ \bibnamefont {Dean}},\ }\href
  {https://science.sciencemag.org/content/363/6431/1059} {\bibfield  {journal}
  {\bibinfo  {journal} {Science}\ }\textbf {\bibinfo {volume} {363}},\ \bibinfo
  {pages} {1059} (\bibinfo {year} {2019})}\BibitemShut {NoStop}%
\bibitem [{\citenamefont {Wu}\ \emph {et~al.}(2018)\citenamefont {Wu},
  \citenamefont {MacDonald},\ and\ \citenamefont
  {Martin}}]{PhysRevLett.121.257001}%
  \BibitemOpen
  \bibfield  {author} {\bibinfo {author} {\bibfnamefont {F.}~\bibnamefont
  {Wu}}, \bibinfo {author} {\bibfnamefont {A.~H.}\ \bibnamefont {MacDonald}}, \
  and\ \bibinfo {author} {\bibfnamefont {I.}~\bibnamefont {Martin}},\ }\href
  {\doibase 10.1103/PhysRevLett.121.257001} {\bibfield  {journal} {\bibinfo
  {journal} {Phys. Rev. Lett.}\ }\textbf {\bibinfo {volume} {121}},\ \bibinfo
  {pages} {257001} (\bibinfo {year} {2018})}\BibitemShut {NoStop}%
\bibitem [{\citenamefont {Lian}\ \emph {et~al.}(2019)\citenamefont {Lian},
  \citenamefont {Wang},\ and\ \citenamefont
  {Bernevig}}]{PhysRevLett.122.257002}%
  \BibitemOpen
  \bibfield  {author} {\bibinfo {author} {\bibfnamefont {B.}~\bibnamefont
  {Lian}}, \bibinfo {author} {\bibfnamefont {Z.}~\bibnamefont {Wang}}, \ and\
  \bibinfo {author} {\bibfnamefont {B.~A.}\ \bibnamefont {Bernevig}},\ }\href
  {\doibase 10.1103/PhysRevLett.122.257002} {\bibfield  {journal} {\bibinfo
  {journal} {Phys. Rev. Lett.}\ }\textbf {\bibinfo {volume} {122}},\ \bibinfo
  {pages} {257002} (\bibinfo {year} {2019})}\BibitemShut {NoStop}%
\bibitem [{\citenamefont {Zupancic}\ \emph {et~al.}(2016)\citenamefont
  {Zupancic}, \citenamefont {Preiss}, \citenamefont {Ma}, \citenamefont
  {Lukin}, \citenamefont {Tai}, \citenamefont {Rispoli}, \citenamefont
  {Islam},\ and\ \citenamefont {Greiner}}]{zupancic2016ultra}%
  \BibitemOpen
  \bibfield  {author} {\bibinfo {author} {\bibfnamefont {P.}~\bibnamefont
  {Zupancic}}, \bibinfo {author} {\bibfnamefont {P.~M.}\ \bibnamefont
  {Preiss}}, \bibinfo {author} {\bibfnamefont {R.}~\bibnamefont {Ma}}, \bibinfo
  {author} {\bibfnamefont {A.}~\bibnamefont {Lukin}}, \bibinfo {author}
  {\bibfnamefont {M.~E.}\ \bibnamefont {Tai}}, \bibinfo {author} {\bibfnamefont
  {M.}~\bibnamefont {Rispoli}}, \bibinfo {author} {\bibfnamefont
  {R.}~\bibnamefont {Islam}}, \ and\ \bibinfo {author} {\bibfnamefont
  {M.}~\bibnamefont {Greiner}},\ }\href
  {https://www.osapublishing.org/oe/abstract.cfm?uri=oe-24-13-13881} {\bibfield
   {journal} {\bibinfo  {journal} {Optics express}\ }\textbf {\bibinfo {volume}
  {24}},\ \bibinfo {pages} {13881} (\bibinfo {year} {2016})}\BibitemShut
  {NoStop}%
\bibitem [{\citenamefont {Barredo}\ \emph {et~al.}(2016)\citenamefont
  {Barredo}, \citenamefont {de~L{\'e}s{\'e}leuc}, \citenamefont {Lienhard},
  \citenamefont {Lahaye},\ and\ \citenamefont {Browaeys}}]{barredo2016atom}%
  \BibitemOpen
  \bibfield  {author} {\bibinfo {author} {\bibfnamefont {D.}~\bibnamefont
  {Barredo}}, \bibinfo {author} {\bibfnamefont {S.}~\bibnamefont
  {de~L{\'e}s{\'e}leuc}}, \bibinfo {author} {\bibfnamefont {V.}~\bibnamefont
  {Lienhard}}, \bibinfo {author} {\bibfnamefont {T.}~\bibnamefont {Lahaye}}, \
  and\ \bibinfo {author} {\bibfnamefont {A.}~\bibnamefont {Browaeys}},\ }\href
  {http://science.sciencemag.org/content/354/6315/1021} {\bibfield  {journal}
  {\bibinfo  {journal} {Science}\ }\textbf {\bibinfo {volume} {354}},\ \bibinfo
  {pages} {1021} (\bibinfo {year} {2016})}\BibitemShut {NoStop}%
\bibitem [{\citenamefont {Endres}\ \emph {et~al.}(2016)\citenamefont {Endres},
  \citenamefont {Bernien}, \citenamefont {Keesling}, \citenamefont {Levine},
  \citenamefont {Anschuetz}, \citenamefont {Krajenbrink}, \citenamefont
  {Senko}, \citenamefont {Vuletic}, \citenamefont {Greiner},\ and\
  \citenamefont {Lukin}}]{endres2016atom}%
  \BibitemOpen
  \bibfield  {author} {\bibinfo {author} {\bibfnamefont {M.}~\bibnamefont
  {Endres}}, \bibinfo {author} {\bibfnamefont {H.}~\bibnamefont {Bernien}},
  \bibinfo {author} {\bibfnamefont {A.}~\bibnamefont {Keesling}}, \bibinfo
  {author} {\bibfnamefont {H.}~\bibnamefont {Levine}}, \bibinfo {author}
  {\bibfnamefont {E.~R.}\ \bibnamefont {Anschuetz}}, \bibinfo {author}
  {\bibfnamefont {A.}~\bibnamefont {Krajenbrink}}, \bibinfo {author}
  {\bibfnamefont {C.}~\bibnamefont {Senko}}, \bibinfo {author} {\bibfnamefont
  {V.}~\bibnamefont {Vuletic}}, \bibinfo {author} {\bibfnamefont
  {M.}~\bibnamefont {Greiner}}, \ and\ \bibinfo {author} {\bibfnamefont
  {M.~D.}\ \bibnamefont {Lukin}},\ }\href
  {http://science.sciencemag.org/content/354/6315/1024} {\bibfield  {journal}
  {\bibinfo  {journal} {Science}\ }\textbf {\bibinfo {volume} {354}},\ \bibinfo
  {pages} {1024} (\bibinfo {year} {2016})}\BibitemShut {NoStop}%
\bibitem [{\citenamefont {Barredo}\ \emph {et~al.}(2018)\citenamefont
  {Barredo}, \citenamefont {Lienhard}, \citenamefont {De~Leseleuc},
  \citenamefont {Lahaye},\ and\ \citenamefont
  {Browaeys}}]{barredo2018synthetic}%
  \BibitemOpen
  \bibfield  {author} {\bibinfo {author} {\bibfnamefont {D.}~\bibnamefont
  {Barredo}}, \bibinfo {author} {\bibfnamefont {V.}~\bibnamefont {Lienhard}},
  \bibinfo {author} {\bibfnamefont {S.}~\bibnamefont {De~Leseleuc}}, \bibinfo
  {author} {\bibfnamefont {T.}~\bibnamefont {Lahaye}}, \ and\ \bibinfo {author}
  {\bibfnamefont {A.}~\bibnamefont {Browaeys}},\ }\href
  {https://www.nature.com/articles/s41586-018-0450-2} {\bibfield  {journal}
  {\bibinfo  {journal} {Nature}\ }\textbf {\bibinfo {volume} {561}},\ \bibinfo
  {pages} {79} (\bibinfo {year} {2018})}\BibitemShut {NoStop}%
\bibitem [{\citenamefont {Schine}\ \emph {et~al.}(2019)\citenamefont {Schine},
  \citenamefont {Chalupnik}, \citenamefont {Can}, \citenamefont {Gromov},\ and\
  \citenamefont {Simon}}]{schine2019electromagnetic}%
  \BibitemOpen
  \bibfield  {author} {\bibinfo {author} {\bibfnamefont {N.}~\bibnamefont
  {Schine}}, \bibinfo {author} {\bibfnamefont {M.}~\bibnamefont {Chalupnik}},
  \bibinfo {author} {\bibfnamefont {T.}~\bibnamefont {Can}}, \bibinfo {author}
  {\bibfnamefont {A.}~\bibnamefont {Gromov}}, \ and\ \bibinfo {author}
  {\bibfnamefont {J.}~\bibnamefont {Simon}},\ }\href
  {https://www.nature.com/articles/s41586-018-0817-4} {\bibfield  {journal}
  {\bibinfo  {journal} {Nature}\ }\textbf {\bibinfo {volume} {565}},\ \bibinfo
  {pages} {173} (\bibinfo {year} {2019})}\BibitemShut {NoStop}%
\bibitem [{\citenamefont {Fazal}\ and\ \citenamefont
  {Block}(2011)}]{fazal2011optical}%
  \BibitemOpen
  \bibfield  {author} {\bibinfo {author} {\bibfnamefont {F.~M.}\ \bibnamefont
  {Fazal}}\ and\ \bibinfo {author} {\bibfnamefont {S.~M.}\ \bibnamefont
  {Block}},\ }\href {https://www.nature.com/articles/nphoton.2011.100}
  {\bibfield  {journal} {\bibinfo  {journal} {Nature photonics}\ }\textbf
  {\bibinfo {volume} {5}},\ \bibinfo {pages} {318} (\bibinfo {year}
  {2011})}\BibitemShut {NoStop}%
\bibitem [{\citenamefont {Tai}\ \emph {et~al.}(2017)\citenamefont {Tai},
  \citenamefont {Lukin}, \citenamefont {Rispoli}, \citenamefont {Schittko},
  \citenamefont {Menke}, \citenamefont {Borgnia}, \citenamefont {Preiss},
  \citenamefont {Grusdt}, \citenamefont {Kaufman},\ and\ \citenamefont
  {Greiner}}]{tai2017microscopy}%
  \BibitemOpen
  \bibfield  {author} {\bibinfo {author} {\bibfnamefont {M.~E.}\ \bibnamefont
  {Tai}}, \bibinfo {author} {\bibfnamefont {A.}~\bibnamefont {Lukin}}, \bibinfo
  {author} {\bibfnamefont {M.}~\bibnamefont {Rispoli}}, \bibinfo {author}
  {\bibfnamefont {R.}~\bibnamefont {Schittko}}, \bibinfo {author}
  {\bibfnamefont {T.}~\bibnamefont {Menke}}, \bibinfo {author} {\bibfnamefont
  {D.}~\bibnamefont {Borgnia}}, \bibinfo {author} {\bibfnamefont {P.~M.}\
  \bibnamefont {Preiss}}, \bibinfo {author} {\bibfnamefont {F.}~\bibnamefont
  {Grusdt}}, \bibinfo {author} {\bibfnamefont {A.~M.}\ \bibnamefont {Kaufman}},
  \ and\ \bibinfo {author} {\bibfnamefont {M.}~\bibnamefont {Greiner}},\ }\href
  {https://www.nature.com/articles/nature22811} {\bibfield  {journal} {\bibinfo
   {journal} {Nature}\ }\textbf {\bibinfo {volume} {546}},\ \bibinfo {pages}
  {519} (\bibinfo {year} {2017})}\BibitemShut {NoStop}%
\bibitem [{\citenamefont {Lukin}\ \emph {et~al.}(2019)\citenamefont {Lukin},
  \citenamefont {Rispoli}, \citenamefont {Schittko}, \citenamefont {Tai},
  \citenamefont {Kaufman}, \citenamefont {Choi}, \citenamefont {Khemani},
  \citenamefont {L{\'e}onard},\ and\ \citenamefont
  {Greiner}}]{lukin2019probing}%
  \BibitemOpen
  \bibfield  {author} {\bibinfo {author} {\bibfnamefont {A.}~\bibnamefont
  {Lukin}}, \bibinfo {author} {\bibfnamefont {M.}~\bibnamefont {Rispoli}},
  \bibinfo {author} {\bibfnamefont {R.}~\bibnamefont {Schittko}}, \bibinfo
  {author} {\bibfnamefont {M.~E.}\ \bibnamefont {Tai}}, \bibinfo {author}
  {\bibfnamefont {A.~M.}\ \bibnamefont {Kaufman}}, \bibinfo {author}
  {\bibfnamefont {S.}~\bibnamefont {Choi}}, \bibinfo {author} {\bibfnamefont
  {V.}~\bibnamefont {Khemani}}, \bibinfo {author} {\bibfnamefont
  {J.}~\bibnamefont {L{\'e}onard}}, \ and\ \bibinfo {author} {\bibfnamefont
  {M.}~\bibnamefont {Greiner}},\ }\href
  {https://science.sciencemag.org/content/364/6437/256} {\bibfield  {journal}
  {\bibinfo  {journal} {Science}\ }\textbf {\bibinfo {volume} {364}},\ \bibinfo
  {pages} {256} (\bibinfo {year} {2019})}\BibitemShut {NoStop}%
\bibitem [{\citenamefont {Chiu}\ \emph {et~al.}(2019)\citenamefont {Chiu},
  \citenamefont {Ji}, \citenamefont {Bohrdt}, \citenamefont {Xu}, \citenamefont
  {Knap}, \citenamefont {Demler}, \citenamefont {Grusdt}, \citenamefont
  {Greiner},\ and\ \citenamefont {Greif}}]{chiu2019string}%
  \BibitemOpen
  \bibfield  {author} {\bibinfo {author} {\bibfnamefont {C.~S.}\ \bibnamefont
  {Chiu}}, \bibinfo {author} {\bibfnamefont {G.}~\bibnamefont {Ji}}, \bibinfo
  {author} {\bibfnamefont {A.}~\bibnamefont {Bohrdt}}, \bibinfo {author}
  {\bibfnamefont {M.}~\bibnamefont {Xu}}, \bibinfo {author} {\bibfnamefont
  {M.}~\bibnamefont {Knap}}, \bibinfo {author} {\bibfnamefont {E.}~\bibnamefont
  {Demler}}, \bibinfo {author} {\bibfnamefont {F.}~\bibnamefont {Grusdt}},
  \bibinfo {author} {\bibfnamefont {M.}~\bibnamefont {Greiner}}, \ and\
  \bibinfo {author} {\bibfnamefont {D.}~\bibnamefont {Greif}},\ }\href
  {https://science.sciencemag.org/content/365/6450/251} {\bibfield  {journal}
  {\bibinfo  {journal} {Science}\ }\textbf {\bibinfo {volume} {365}},\ \bibinfo
  {pages} {251} (\bibinfo {year} {2019})}\BibitemShut {NoStop}%
\bibitem [{\citenamefont {Covey}\ \emph {et~al.}(2019)\citenamefont {Covey},
  \citenamefont {Madjarov}, \citenamefont {Cooper},\ and\ \citenamefont
  {Endres}}]{PhysRevLett.122.173201}%
  \BibitemOpen
  \bibfield  {author} {\bibinfo {author} {\bibfnamefont {J.~P.}\ \bibnamefont
  {Covey}}, \bibinfo {author} {\bibfnamefont {I.~S.}\ \bibnamefont {Madjarov}},
  \bibinfo {author} {\bibfnamefont {A.}~\bibnamefont {Cooper}}, \ and\ \bibinfo
  {author} {\bibfnamefont {M.}~\bibnamefont {Endres}},\ }\href {\doibase
  10.1103/PhysRevLett.122.173201} {\bibfield  {journal} {\bibinfo  {journal}
  {Phys. Rev. Lett.}\ }\textbf {\bibinfo {volume} {122}},\ \bibinfo {pages}
  {173201} (\bibinfo {year} {2019})}\BibitemShut {NoStop}%
\bibitem [{\citenamefont {Madjarov}\ \emph {et~al.}(2019)\citenamefont
  {Madjarov}, \citenamefont {Cooper}, \citenamefont {Shaw}, \citenamefont
  {Covey}, \citenamefont {Schkolnik}, \citenamefont {Yoon}, \citenamefont
  {Williams},\ and\ \citenamefont {Endres}}]{PhysRevX.9.041052}%
  \BibitemOpen
  \bibfield  {author} {\bibinfo {author} {\bibfnamefont {I.~S.}\ \bibnamefont
  {Madjarov}}, \bibinfo {author} {\bibfnamefont {A.}~\bibnamefont {Cooper}},
  \bibinfo {author} {\bibfnamefont {A.~L.}\ \bibnamefont {Shaw}}, \bibinfo
  {author} {\bibfnamefont {J.~P.}\ \bibnamefont {Covey}}, \bibinfo {author}
  {\bibfnamefont {V.}~\bibnamefont {Schkolnik}}, \bibinfo {author}
  {\bibfnamefont {T.~H.}\ \bibnamefont {Yoon}}, \bibinfo {author}
  {\bibfnamefont {J.~R.}\ \bibnamefont {Williams}}, \ and\ \bibinfo {author}
  {\bibfnamefont {M.}~\bibnamefont {Endres}},\ }\href {\doibase
  10.1103/PhysRevX.9.041052} {\bibfield  {journal} {\bibinfo  {journal} {Phys.
  Rev. X}\ }\textbf {\bibinfo {volume} {9}},\ \bibinfo {pages} {041052}
  (\bibinfo {year} {2019})}\BibitemShut {NoStop}%
\bibitem [{\citenamefont {Oka}\ and\ \citenamefont
  {Aoki}(2009)}]{PhysRevB.79.081406}%
  \BibitemOpen
  \bibfield  {author} {\bibinfo {author} {\bibfnamefont {T.}~\bibnamefont
  {Oka}}\ and\ \bibinfo {author} {\bibfnamefont {H.}~\bibnamefont {Aoki}},\
  }\href {\doibase 10.1103/PhysRevB.79.081406} {\bibfield  {journal} {\bibinfo
  {journal} {Phys. Rev. B}\ }\textbf {\bibinfo {volume} {79}},\ \bibinfo
  {pages} {081406} (\bibinfo {year} {2009})}\BibitemShut {NoStop}%
\bibitem [{\citenamefont {Kitagawa}\ \emph {et~al.}(2011)\citenamefont
  {Kitagawa}, \citenamefont {Oka}, \citenamefont {Brataas}, \citenamefont
  {Fu},\ and\ \citenamefont {Demler}}]{kitagawa2011transport}%
  \BibitemOpen
  \bibfield  {author} {\bibinfo {author} {\bibfnamefont {T.}~\bibnamefont
  {Kitagawa}}, \bibinfo {author} {\bibfnamefont {T.}~\bibnamefont {Oka}},
  \bibinfo {author} {\bibfnamefont {A.}~\bibnamefont {Brataas}}, \bibinfo
  {author} {\bibfnamefont {L.}~\bibnamefont {Fu}}, \ and\ \bibinfo {author}
  {\bibfnamefont {E.}~\bibnamefont {Demler}},\ }\href@noop {} {\bibfield
  {journal} {\bibinfo  {journal} {Physical Review B}\ }\textbf {\bibinfo
  {volume} {84}},\ \bibinfo {pages} {235108} (\bibinfo {year}
  {2011})}\BibitemShut {NoStop}%
\bibitem [{\citenamefont {Lindner}\ \emph {et~al.}(2011)\citenamefont
  {Lindner}, \citenamefont {Refael},\ and\ \citenamefont
  {Galitski}}]{lindner2011floquet}%
  \BibitemOpen
  \bibfield  {author} {\bibinfo {author} {\bibfnamefont {N.~H.}\ \bibnamefont
  {Lindner}}, \bibinfo {author} {\bibfnamefont {G.}~\bibnamefont {Refael}}, \
  and\ \bibinfo {author} {\bibfnamefont {V.}~\bibnamefont {Galitski}},\ }\href
  {https://www.nature.com/articles/nphys1926} {\bibfield  {journal} {\bibinfo
  {journal} {Nature Physics}\ }\textbf {\bibinfo {volume} {7}},\ \bibinfo
  {pages} {490} (\bibinfo {year} {2011})}\BibitemShut {NoStop}%
\bibitem [{\citenamefont {Wang}\ \emph {et~al.}(2013)\citenamefont {Wang},
  \citenamefont {Steinberg}, \citenamefont {Jarillo-Herrero},\ and\
  \citenamefont {Gedik}}]{wang2013observation}%
  \BibitemOpen
  \bibfield  {author} {\bibinfo {author} {\bibfnamefont {Y.}~\bibnamefont
  {Wang}}, \bibinfo {author} {\bibfnamefont {H.}~\bibnamefont {Steinberg}},
  \bibinfo {author} {\bibfnamefont {P.}~\bibnamefont {Jarillo-Herrero}}, \ and\
  \bibinfo {author} {\bibfnamefont {N.}~\bibnamefont {Gedik}},\ }\href@noop {}
  {\bibfield  {journal} {\bibinfo  {journal} {Science}\ }\textbf {\bibinfo
  {volume} {342}},\ \bibinfo {pages} {453} (\bibinfo {year}
  {2013})}\BibitemShut {NoStop}%
\bibitem [{\citenamefont {McIver}\ \emph {et~al.}(2020)\citenamefont {McIver},
  \citenamefont {Schulte}, \citenamefont {Stein}, \citenamefont {Matsuyama},
  \citenamefont {Jotzu}, \citenamefont {Meier},\ and\ \citenamefont
  {Cavalleri}}]{mciver2019light}%
  \BibitemOpen
  \bibfield  {author} {\bibinfo {author} {\bibfnamefont {J.~W.}\ \bibnamefont
  {McIver}}, \bibinfo {author} {\bibfnamefont {B.}~\bibnamefont {Schulte}},
  \bibinfo {author} {\bibfnamefont {F.-U.}\ \bibnamefont {Stein}}, \bibinfo
  {author} {\bibfnamefont {T.}~\bibnamefont {Matsuyama}}, \bibinfo {author}
  {\bibfnamefont {G.}~\bibnamefont {Jotzu}}, \bibinfo {author} {\bibfnamefont
  {G.}~\bibnamefont {Meier}}, \ and\ \bibinfo {author} {\bibfnamefont
  {A.}~\bibnamefont {Cavalleri}},\ }\href
  {https://www.nature.com/articles/s41567-019-0698-y} {\bibfield  {journal}
  {\bibinfo  {journal} {Nature Physics}\ }\textbf {\bibinfo {volume} {16}},\
  \bibinfo {pages} {38} (\bibinfo {year} {2020})}\BibitemShut {NoStop}%
\bibitem [{\citenamefont {Usaj}\ \emph {et~al.}(2014)\citenamefont {Usaj},
  \citenamefont {Perez-Piskunow}, \citenamefont {Foa~Torres},\ and\
  \citenamefont {Balseiro}}]{PhysRevB.90.115423}%
  \BibitemOpen
  \bibfield  {author} {\bibinfo {author} {\bibfnamefont {G.}~\bibnamefont
  {Usaj}}, \bibinfo {author} {\bibfnamefont {P.~M.}\ \bibnamefont
  {Perez-Piskunow}}, \bibinfo {author} {\bibfnamefont {L.~E.~F.}\ \bibnamefont
  {Foa~Torres}}, \ and\ \bibinfo {author} {\bibfnamefont {C.~A.}\ \bibnamefont
  {Balseiro}},\ }\href {\doibase 10.1103/PhysRevB.90.115423} {\bibfield
  {journal} {\bibinfo  {journal} {Phys. Rev. B}\ }\textbf {\bibinfo {volume}
  {90}},\ \bibinfo {pages} {115423} (\bibinfo {year} {2014})}\BibitemShut
  {NoStop}%
\bibitem [{\citenamefont {Topp}\ \emph {et~al.}(2019)\citenamefont {Topp},
  \citenamefont {Jotzu}, \citenamefont {McIver}, \citenamefont {Xian},
  \citenamefont {Rubio},\ and\ \citenamefont
  {Sentef}}]{PhysRevResearch.1.023031}%
  \BibitemOpen
  \bibfield  {author} {\bibinfo {author} {\bibfnamefont {G.~E.}\ \bibnamefont
  {Topp}}, \bibinfo {author} {\bibfnamefont {G.}~\bibnamefont {Jotzu}},
  \bibinfo {author} {\bibfnamefont {J.~W.}\ \bibnamefont {McIver}}, \bibinfo
  {author} {\bibfnamefont {L.}~\bibnamefont {Xian}}, \bibinfo {author}
  {\bibfnamefont {A.}~\bibnamefont {Rubio}}, \ and\ \bibinfo {author}
  {\bibfnamefont {M.~A.}\ \bibnamefont {Sentef}},\ }\href {\doibase
  10.1103/PhysRevResearch.1.023031} {\bibfield  {journal} {\bibinfo  {journal}
  {Phys. Rev. Research}\ }\textbf {\bibinfo {volume} {1}},\ \bibinfo {pages}
  {023031} (\bibinfo {year} {2019})}\BibitemShut {NoStop}%
\bibitem [{\citenamefont {Li}\ \emph {et~al.}(2019)\citenamefont {Li},
  \citenamefont {Fertig},\ and\ \citenamefont {Seradjeh}}]{li2019floquet}%
  \BibitemOpen
  \bibfield  {author} {\bibinfo {author} {\bibfnamefont {Y.}~\bibnamefont
  {Li}}, \bibinfo {author} {\bibfnamefont {H.}~\bibnamefont {Fertig}}, \ and\
  \bibinfo {author} {\bibfnamefont {B.}~\bibnamefont {Seradjeh}},\ }\href
  {https://arxiv.org/abs/1910.04711} {\bibfield  {journal} {\bibinfo  {journal}
  {arXiv preprint arXiv:1910.04711}\ } (\bibinfo {year} {2019})}\BibitemShut
  {NoStop}%
\bibitem [{\citenamefont {Katz}\ \emph {et~al.}(2019)\citenamefont {Katz},
  \citenamefont {Refael},\ and\ \citenamefont {Lindner}}]{katz2019optically}%
  \BibitemOpen
  \bibfield  {author} {\bibinfo {author} {\bibfnamefont {O.}~\bibnamefont
  {Katz}}, \bibinfo {author} {\bibfnamefont {G.}~\bibnamefont {Refael}}, \ and\
  \bibinfo {author} {\bibfnamefont {N.~H.}\ \bibnamefont {Lindner}},\ }\href
  {https://arxiv.org/abs/1910.13510} {\bibfield  {journal} {\bibinfo  {journal}
  {arXiv}\ ,\ \bibinfo {pages} {arXiv}} (\bibinfo {year} {2019})}\BibitemShut
  {NoStop}%
\bibitem [{\citenamefont {Karch}\ \emph {et~al.}(2011)\citenamefont {Karch},
  \citenamefont {Drexler}, \citenamefont {Olbrich}, \citenamefont
  {Fehrenbacher}, \citenamefont {Hirmer}, \citenamefont {Glazov}, \citenamefont
  {Tarasenko}, \citenamefont {Ivchenko}, \citenamefont {Birkner}, \citenamefont
  {Eroms}, \citenamefont {Weiss}, \citenamefont {Yakimova}, \citenamefont
  {Lara-Avila}, \citenamefont {Kubatkin}, \citenamefont {Ostler}, \citenamefont
  {Seyller},\ and\ \citenamefont {Ganichev}}]{PhysRevLett.107.276601}%
  \BibitemOpen
  \bibfield  {author} {\bibinfo {author} {\bibfnamefont {J.}~\bibnamefont
  {Karch}}, \bibinfo {author} {\bibfnamefont {C.}~\bibnamefont {Drexler}},
  \bibinfo {author} {\bibfnamefont {P.}~\bibnamefont {Olbrich}}, \bibinfo
  {author} {\bibfnamefont {M.}~\bibnamefont {Fehrenbacher}}, \bibinfo {author}
  {\bibfnamefont {M.}~\bibnamefont {Hirmer}}, \bibinfo {author} {\bibfnamefont
  {M.~M.}\ \bibnamefont {Glazov}}, \bibinfo {author} {\bibfnamefont {S.~A.}\
  \bibnamefont {Tarasenko}}, \bibinfo {author} {\bibfnamefont {E.~L.}\
  \bibnamefont {Ivchenko}}, \bibinfo {author} {\bibfnamefont {B.}~\bibnamefont
  {Birkner}}, \bibinfo {author} {\bibfnamefont {J.}~\bibnamefont {Eroms}},
  \bibinfo {author} {\bibfnamefont {D.}~\bibnamefont {Weiss}}, \bibinfo
  {author} {\bibfnamefont {R.}~\bibnamefont {Yakimova}}, \bibinfo {author}
  {\bibfnamefont {S.}~\bibnamefont {Lara-Avila}}, \bibinfo {author}
  {\bibfnamefont {S.}~\bibnamefont {Kubatkin}}, \bibinfo {author}
  {\bibfnamefont {M.}~\bibnamefont {Ostler}}, \bibinfo {author} {\bibfnamefont
  {T.}~\bibnamefont {Seyller}}, \ and\ \bibinfo {author} {\bibfnamefont
  {S.~D.}\ \bibnamefont {Ganichev}},\ }\href {\doibase
  10.1103/PhysRevLett.107.276601} {\bibfield  {journal} {\bibinfo  {journal}
  {Phys. Rev. Lett.}\ }\textbf {\bibinfo {volume} {107}},\ \bibinfo {pages}
  {276601} (\bibinfo {year} {2011})}\BibitemShut {NoStop}%
\bibitem [{\citenamefont {Castro~Neto}\ \emph {et~al.}(2009)\citenamefont
  {Castro~Neto}, \citenamefont {Guinea}, \citenamefont {Peres}, \citenamefont
  {Novoselov},\ and\ \citenamefont {Geim}}]{RevModPhys.81.109}%
  \BibitemOpen
  \bibfield  {author} {\bibinfo {author} {\bibfnamefont {A.~H.}\ \bibnamefont
  {Castro~Neto}}, \bibinfo {author} {\bibfnamefont {F.}~\bibnamefont {Guinea}},
  \bibinfo {author} {\bibfnamefont {N.~M.~R.}\ \bibnamefont {Peres}}, \bibinfo
  {author} {\bibfnamefont {K.~S.}\ \bibnamefont {Novoselov}}, \ and\ \bibinfo
  {author} {\bibfnamefont {A.~K.}\ \bibnamefont {Geim}},\ }\href {\doibase
  10.1103/RevModPhys.81.109} {\bibfield  {journal} {\bibinfo  {journal} {Rev.
  Mod. Phys.}\ }\textbf {\bibinfo {volume} {81}},\ \bibinfo {pages} {109}
  (\bibinfo {year} {2009})}\BibitemShut {NoStop}%
\bibitem [{\citenamefont {G\'omez-Le\'on}\ and\ \citenamefont
  {Platero}(2013)}]{PhysRevLett.110.200403}%
  \BibitemOpen
  \bibfield  {author} {\bibinfo {author} {\bibfnamefont {A.}~\bibnamefont
  {G\'omez-Le\'on}}\ and\ \bibinfo {author} {\bibfnamefont {G.}~\bibnamefont
  {Platero}},\ }\href {\doibase 10.1103/PhysRevLett.110.200403} {\bibfield
  {journal} {\bibinfo  {journal} {Phys. Rev. Lett.}\ }\textbf {\bibinfo
  {volume} {110}},\ \bibinfo {pages} {200403} (\bibinfo {year}
  {2013})}\BibitemShut {NoStop}%
\bibitem [{\citenamefont {Goldman}\ and\ \citenamefont
  {Dalibard}(2014)}]{PhysRevX.4.031027}%
  \BibitemOpen
  \bibfield  {author} {\bibinfo {author} {\bibfnamefont {N.}~\bibnamefont
  {Goldman}}\ and\ \bibinfo {author} {\bibfnamefont {J.}~\bibnamefont
  {Dalibard}},\ }\href {\doibase 10.1103/PhysRevX.4.031027} {\bibfield
  {journal} {\bibinfo  {journal} {Phys. Rev. X}\ }\textbf {\bibinfo {volume}
  {4}},\ \bibinfo {pages} {031027} (\bibinfo {year} {2014})}\BibitemShut
  {NoStop}%
\bibitem [{\citenamefont {Dehghani}\ \emph {et~al.}(2014)\citenamefont
  {Dehghani}, \citenamefont {Oka},\ and\ \citenamefont
  {Mitra}}]{HDehghani2014Dissipative}%
  \BibitemOpen
  \bibfield  {author} {\bibinfo {author} {\bibfnamefont {H.}~\bibnamefont
  {Dehghani}}, \bibinfo {author} {\bibfnamefont {T.}~\bibnamefont {Oka}}, \
  and\ \bibinfo {author} {\bibfnamefont {A.}~\bibnamefont {Mitra}},\ }\href
  {\doibase 10.1103/PhysRevB.90.195429} {\bibfield  {journal} {\bibinfo
  {journal} {Phys. Rev. B}\ }\textbf {\bibinfo {volume} {90}},\ \bibinfo
  {pages} {195429} (\bibinfo {year} {2014})}\BibitemShut {NoStop}%
\bibitem [{\citenamefont {Dehghani}\ \emph {et~al.}(2015)\citenamefont
  {Dehghani}, \citenamefont {Oka},\ and\ \citenamefont
  {Mitra}}]{HDehghani2015Out}%
  \BibitemOpen
  \bibfield  {author} {\bibinfo {author} {\bibfnamefont {H.}~\bibnamefont
  {Dehghani}}, \bibinfo {author} {\bibfnamefont {T.}~\bibnamefont {Oka}}, \
  and\ \bibinfo {author} {\bibfnamefont {A.}~\bibnamefont {Mitra}},\ }\href
  {\doibase 10.1103/PhysRevB.91.155422} {\bibfield  {journal} {\bibinfo
  {journal} {Phys. Rev. B}\ }\textbf {\bibinfo {volume} {91}},\ \bibinfo
  {pages} {155422} (\bibinfo {year} {2015})}\BibitemShut {NoStop}%
\bibitem [{\citenamefont {Seetharam}\ \emph {et~al.}(2019)\citenamefont
  {Seetharam}, \citenamefont {Bardyn}, \citenamefont {Lindner}, \citenamefont
  {Rudner},\ and\ \citenamefont {Refael}}]{PhysRevB.99.014307}%
  \BibitemOpen
  \bibfield  {author} {\bibinfo {author} {\bibfnamefont {K.~I.}\ \bibnamefont
  {Seetharam}}, \bibinfo {author} {\bibfnamefont {C.-E.}\ \bibnamefont
  {Bardyn}}, \bibinfo {author} {\bibfnamefont {N.~H.}\ \bibnamefont {Lindner}},
  \bibinfo {author} {\bibfnamefont {M.~S.}\ \bibnamefont {Rudner}}, \ and\
  \bibinfo {author} {\bibfnamefont {G.}~\bibnamefont {Refael}},\ }\href
  {\doibase 10.1103/PhysRevB.99.014307} {\bibfield  {journal} {\bibinfo
  {journal} {Phys. Rev. B}\ }\textbf {\bibinfo {volume} {99}},\ \bibinfo
  {pages} {014307} (\bibinfo {year} {2019})}\BibitemShut {NoStop}%
\bibitem [{\citenamefont {Eckardt}\ and\ \citenamefont
  {Anisimovas}(2015)}]{eckardt2015high}%
  \BibitemOpen
  \bibfield  {author} {\bibinfo {author} {\bibfnamefont {A.}~\bibnamefont
  {Eckardt}}\ and\ \bibinfo {author} {\bibfnamefont {E.}~\bibnamefont
  {Anisimovas}},\ }\href
  {https://iopscience.iop.org/article/10.1088/1367-2630/17/9/093039} {\bibfield
   {journal} {\bibinfo  {journal} {New journal of physics}\ }\textbf {\bibinfo
  {volume} {17}},\ \bibinfo {pages} {093039} (\bibinfo {year}
  {2015})}\BibitemShut {NoStop}%
\bibitem [{\citenamefont {Mikami}\ \emph {et~al.}(2016)\citenamefont {Mikami},
  \citenamefont {Kitamura}, \citenamefont {Yasuda}, \citenamefont {Tsuji},
  \citenamefont {Oka},\ and\ \citenamefont {Aoki}}]{PhysRevB.93.144307}%
  \BibitemOpen
  \bibfield  {author} {\bibinfo {author} {\bibfnamefont {T.}~\bibnamefont
  {Mikami}}, \bibinfo {author} {\bibfnamefont {S.}~\bibnamefont {Kitamura}},
  \bibinfo {author} {\bibfnamefont {K.}~\bibnamefont {Yasuda}}, \bibinfo
  {author} {\bibfnamefont {N.}~\bibnamefont {Tsuji}}, \bibinfo {author}
  {\bibfnamefont {T.}~\bibnamefont {Oka}}, \ and\ \bibinfo {author}
  {\bibfnamefont {H.}~\bibnamefont {Aoki}},\ }\href {\doibase
  10.1103/PhysRevB.93.144307} {\bibfield  {journal} {\bibinfo  {journal} {Phys.
  Rev. B}\ }\textbf {\bibinfo {volume} {93}},\ \bibinfo {pages} {144307}
  (\bibinfo {year} {2016})}\BibitemShut {NoStop}%
\bibitem [{\citenamefont {Katan}\ and\ \citenamefont
  {Podolsky}(2013)}]{PhysRevLett.110.016802}%
  \BibitemOpen
  \bibfield  {author} {\bibinfo {author} {\bibfnamefont {Y.~T.}\ \bibnamefont
  {Katan}}\ and\ \bibinfo {author} {\bibfnamefont {D.}~\bibnamefont
  {Podolsky}},\ }\href {\doibase 10.1103/PhysRevLett.110.016802} {\bibfield
  {journal} {\bibinfo  {journal} {Phys. Rev. Lett.}\ }\textbf {\bibinfo
  {volume} {110}},\ \bibinfo {pages} {016802} (\bibinfo {year}
  {2013})}\BibitemShut {NoStop}%
\bibitem [{\citenamefont {Tenenbaum~Katan}\ and\ \citenamefont
  {Podolsky}(2013)}]{PhysRevB.88.224106}%
  \BibitemOpen
  \bibfield  {author} {\bibinfo {author} {\bibfnamefont {Y.}~\bibnamefont
  {Tenenbaum~Katan}}\ and\ \bibinfo {author} {\bibfnamefont {D.}~\bibnamefont
  {Podolsky}},\ }\href {\doibase 10.1103/PhysRevB.88.224106} {\bibfield
  {journal} {\bibinfo  {journal} {Phys. Rev. B}\ }\textbf {\bibinfo {volume}
  {88}},\ \bibinfo {pages} {224106} (\bibinfo {year} {2013})}\BibitemShut
  {NoStop}%
\bibitem [{\citenamefont {Morina}\ \emph {et~al.}(2018)\citenamefont {Morina},
  \citenamefont {Dini}, \citenamefont {Iorsh},\ and\ \citenamefont
  {Shelykh}}]{morina2018optical}%
  \BibitemOpen
  \bibfield  {author} {\bibinfo {author} {\bibfnamefont {S.}~\bibnamefont
  {Morina}}, \bibinfo {author} {\bibfnamefont {K.}~\bibnamefont {Dini}},
  \bibinfo {author} {\bibfnamefont {I.~V.}\ \bibnamefont {Iorsh}}, \ and\
  \bibinfo {author} {\bibfnamefont {I.~A.}\ \bibnamefont {Shelykh}},\ }\href
  {https://pubs.acs.org/doi/10.1021/acsphotonics.7b01394} {\bibfield  {journal}
  {\bibinfo  {journal} {ACS Photonics}\ }\textbf {\bibinfo {volume} {5}},\
  \bibinfo {pages} {1171} (\bibinfo {year} {2018})}\BibitemShut {NoStop}%
\bibitem [{\citenamefont {Rudner}\ \emph {et~al.}(2013)\citenamefont {Rudner},
  \citenamefont {Lindner}, \citenamefont {Berg},\ and\ \citenamefont
  {Levin}}]{PhysRevX.3.031005}%
  \BibitemOpen
  \bibfield  {author} {\bibinfo {author} {\bibfnamefont {M.~S.}\ \bibnamefont
  {Rudner}}, \bibinfo {author} {\bibfnamefont {N.~H.}\ \bibnamefont {Lindner}},
  \bibinfo {author} {\bibfnamefont {E.}~\bibnamefont {Berg}}, \ and\ \bibinfo
  {author} {\bibfnamefont {M.}~\bibnamefont {Levin}},\ }\href {\doibase
  10.1103/PhysRevX.3.031005} {\bibfield  {journal} {\bibinfo  {journal} {Phys.
  Rev. X}\ }\textbf {\bibinfo {volume} {3}},\ \bibinfo {pages} {031005}
  (\bibinfo {year} {2013})}\BibitemShut {NoStop}%
\bibitem [{\citenamefont {Fukui}\ \emph {et~al.}(2005)\citenamefont {Fukui},
  \citenamefont {Hatsugai},\ and\ \citenamefont {Suzuki}}]{fukui2005chern}%
  \BibitemOpen
  \bibfield  {author} {\bibinfo {author} {\bibfnamefont {T.}~\bibnamefont
  {Fukui}}, \bibinfo {author} {\bibfnamefont {Y.}~\bibnamefont {Hatsugai}}, \
  and\ \bibinfo {author} {\bibfnamefont {H.}~\bibnamefont {Suzuki}},\ }\href
  {https://www.jstage.jst.go.jp/article/jpsj/74/6/74_6_1674/_article/-char/ja/}
  {\bibfield  {journal} {\bibinfo  {journal} {Journal of the Physical Society
  of Japan}\ }\textbf {\bibinfo {volume} {74}},\ \bibinfo {pages} {1674}
  (\bibinfo {year} {2005})}\BibitemShut {NoStop}%
\bibitem [{\citenamefont {Thonhauser}\ \emph {et~al.}(2005)\citenamefont
  {Thonhauser}, \citenamefont {Ceresoli}, \citenamefont {Vanderbilt},\ and\
  \citenamefont {Resta}}]{PhysRevLett.95.137205}%
  \BibitemOpen
  \bibfield  {author} {\bibinfo {author} {\bibfnamefont {T.}~\bibnamefont
  {Thonhauser}}, \bibinfo {author} {\bibfnamefont {D.}~\bibnamefont
  {Ceresoli}}, \bibinfo {author} {\bibfnamefont {D.}~\bibnamefont
  {Vanderbilt}}, \ and\ \bibinfo {author} {\bibfnamefont {R.}~\bibnamefont
  {Resta}},\ }\href {\doibase 10.1103/PhysRevLett.95.137205} {\bibfield
  {journal} {\bibinfo  {journal} {Phys. Rev. Lett.}\ }\textbf {\bibinfo
  {volume} {95}},\ \bibinfo {pages} {137205} (\bibinfo {year}
  {2005})}\BibitemShut {NoStop}%
\bibitem [{\citenamefont {Xiao}\ \emph {et~al.}(2005)\citenamefont {Xiao},
  \citenamefont {Shi},\ and\ \citenamefont {Niu}}]{PhysRevLett.95.137204}%
  \BibitemOpen
  \bibfield  {author} {\bibinfo {author} {\bibfnamefont {D.}~\bibnamefont
  {Xiao}}, \bibinfo {author} {\bibfnamefont {J.}~\bibnamefont {Shi}}, \ and\
  \bibinfo {author} {\bibfnamefont {Q.}~\bibnamefont {Niu}},\ }\href {\doibase
  10.1103/PhysRevLett.95.137204} {\bibfield  {journal} {\bibinfo  {journal}
  {Phys. Rev. Lett.}\ }\textbf {\bibinfo {volume} {95}},\ \bibinfo {pages}
  {137204} (\bibinfo {year} {2005})}\BibitemShut {NoStop}%
\bibitem [{\citenamefont {Shi}\ \emph {et~al.}(2007)\citenamefont {Shi},
  \citenamefont {Vignale}, \citenamefont {Xiao},\ and\ \citenamefont
  {Niu}}]{PhysRevLett.99.197202}%
  \BibitemOpen
  \bibfield  {author} {\bibinfo {author} {\bibfnamefont {J.}~\bibnamefont
  {Shi}}, \bibinfo {author} {\bibfnamefont {G.}~\bibnamefont {Vignale}},
  \bibinfo {author} {\bibfnamefont {D.}~\bibnamefont {Xiao}}, \ and\ \bibinfo
  {author} {\bibfnamefont {Q.}~\bibnamefont {Niu}},\ }\href {\doibase
  10.1103/PhysRevLett.99.197202} {\bibfield  {journal} {\bibinfo  {journal}
  {Phys. Rev. Lett.}\ }\textbf {\bibinfo {volume} {99}},\ \bibinfo {pages}
  {197202} (\bibinfo {year} {2007})}\BibitemShut {NoStop}%
\bibitem [{\citenamefont {Hafezi}\ \emph {et~al.}(2011)\citenamefont {Hafezi},
  \citenamefont {Demler}, \citenamefont {Lukin},\ and\ \citenamefont
  {Taylor}}]{hafezi2011robust}%
  \BibitemOpen
  \bibfield  {author} {\bibinfo {author} {\bibfnamefont {M.}~\bibnamefont
  {Hafezi}}, \bibinfo {author} {\bibfnamefont {E.~A.}\ \bibnamefont {Demler}},
  \bibinfo {author} {\bibfnamefont {M.~D.}\ \bibnamefont {Lukin}}, \ and\
  \bibinfo {author} {\bibfnamefont {J.~M.}\ \bibnamefont {Taylor}},\ }\href
  {https://www.nature.com/articles/nphys2063} {\bibfield  {journal} {\bibinfo
  {journal} {Nature Physics}\ }\textbf {\bibinfo {volume} {7}},\ \bibinfo
  {pages} {907} (\bibinfo {year} {2011})}\BibitemShut {NoStop}%
\bibitem [{\citenamefont {Haldane}(1988)}]{PhysRevLett.61.2015}%
  \BibitemOpen
  \bibfield  {author} {\bibinfo {author} {\bibfnamefont {F.~D.~M.}\
  \bibnamefont {Haldane}},\ }\href {\doibase 10.1103/PhysRevLett.61.2015}
  {\bibfield  {journal} {\bibinfo  {journal} {Phys. Rev. Lett.}\ }\textbf
  {\bibinfo {volume} {61}},\ \bibinfo {pages} {2015} (\bibinfo {year}
  {1988})}\BibitemShut {NoStop}%
\bibitem [{\citenamefont {Jo}\ \emph {et~al.}(2012)\citenamefont {Jo},
  \citenamefont {Guzman}, \citenamefont {Thomas}, \citenamefont {Hosur},
  \citenamefont {Vishwanath},\ and\ \citenamefont
  {Stamper-Kurn}}]{PhysRevLett.108.045305}%
  \BibitemOpen
  \bibfield  {author} {\bibinfo {author} {\bibfnamefont {G.-B.}\ \bibnamefont
  {Jo}}, \bibinfo {author} {\bibfnamefont {J.}~\bibnamefont {Guzman}}, \bibinfo
  {author} {\bibfnamefont {C.~K.}\ \bibnamefont {Thomas}}, \bibinfo {author}
  {\bibfnamefont {P.}~\bibnamefont {Hosur}}, \bibinfo {author} {\bibfnamefont
  {A.}~\bibnamefont {Vishwanath}}, \ and\ \bibinfo {author} {\bibfnamefont
  {D.~M.}\ \bibnamefont {Stamper-Kurn}},\ }\href {\doibase
  10.1103/PhysRevLett.108.045305} {\bibfield  {journal} {\bibinfo  {journal}
  {Phys. Rev. Lett.}\ }\textbf {\bibinfo {volume} {108}},\ \bibinfo {pages}
  {045305} (\bibinfo {year} {2012})}\BibitemShut {NoStop}%
\bibitem [{\citenamefont {Koch}\ \emph {et~al.}(2010)\citenamefont {Koch},
  \citenamefont {Houck}, \citenamefont {Hur},\ and\ \citenamefont
  {Girvin}}]{PhysRevA.82.043811}%
  \BibitemOpen
  \bibfield  {author} {\bibinfo {author} {\bibfnamefont {J.}~\bibnamefont
  {Koch}}, \bibinfo {author} {\bibfnamefont {A.~A.}\ \bibnamefont {Houck}},
  \bibinfo {author} {\bibfnamefont {K.~L.}\ \bibnamefont {Hur}}, \ and\
  \bibinfo {author} {\bibfnamefont {S.~M.}\ \bibnamefont {Girvin}},\ }\href
  {\doibase 10.1103/PhysRevA.82.043811} {\bibfield  {journal} {\bibinfo
  {journal} {Phys. Rev. A}\ }\textbf {\bibinfo {volume} {82}},\ \bibinfo
  {pages} {043811} (\bibinfo {year} {2010})}\BibitemShut {NoStop}%
\bibitem [{\citenamefont {Mahmood}\ \emph {et~al.}(2016)\citenamefont
  {Mahmood}, \citenamefont {Chan}, \citenamefont {Alpichshev}, \citenamefont
  {Gardner}, \citenamefont {Lee}, \citenamefont {Lee},\ and\ \citenamefont
  {Gedik}}]{mahmood2016selective}%
  \BibitemOpen
  \bibfield  {author} {\bibinfo {author} {\bibfnamefont {F.}~\bibnamefont
  {Mahmood}}, \bibinfo {author} {\bibfnamefont {C.-K.}\ \bibnamefont {Chan}},
  \bibinfo {author} {\bibfnamefont {Z.}~\bibnamefont {Alpichshev}}, \bibinfo
  {author} {\bibfnamefont {D.}~\bibnamefont {Gardner}}, \bibinfo {author}
  {\bibfnamefont {Y.}~\bibnamefont {Lee}}, \bibinfo {author} {\bibfnamefont
  {P.~A.}\ \bibnamefont {Lee}}, \ and\ \bibinfo {author} {\bibfnamefont
  {N.}~\bibnamefont {Gedik}},\ }\href {https://doi.org/10.1038/nphys3609}
  {\bibfield  {journal} {\bibinfo  {journal} {Nature Physics}\ }\textbf
  {\bibinfo {volume} {12}},\ \bibinfo {pages} {306} (\bibinfo {year}
  {2016})}\BibitemShut {NoStop}%
\bibitem [{\citenamefont {Levin}\ and\ \citenamefont
  {Stern}(2009)}]{PhysRevLett.103.196803}%
  \BibitemOpen
  \bibfield  {author} {\bibinfo {author} {\bibfnamefont {M.}~\bibnamefont
  {Levin}}\ and\ \bibinfo {author} {\bibfnamefont {A.}~\bibnamefont {Stern}},\
  }\href {\doibase 10.1103/PhysRevLett.103.196803} {\bibfield  {journal}
  {\bibinfo  {journal} {Phys. Rev. Lett.}\ }\textbf {\bibinfo {volume} {103}},\
  \bibinfo {pages} {196803} (\bibinfo {year} {2009})}\BibitemShut {NoStop}%
\bibitem [{\citenamefont {Swingle}\ \emph {et~al.}(2011)\citenamefont
  {Swingle}, \citenamefont {Barkeshli}, \citenamefont {McGreevy},\ and\
  \citenamefont {Senthil}}]{PhysRevB.83.195139}%
  \BibitemOpen
  \bibfield  {author} {\bibinfo {author} {\bibfnamefont {B.}~\bibnamefont
  {Swingle}}, \bibinfo {author} {\bibfnamefont {M.}~\bibnamefont {Barkeshli}},
  \bibinfo {author} {\bibfnamefont {J.}~\bibnamefont {McGreevy}}, \ and\
  \bibinfo {author} {\bibfnamefont {T.}~\bibnamefont {Senthil}},\ }\href
  {\doibase 10.1103/PhysRevB.83.195139} {\bibfield  {journal} {\bibinfo
  {journal} {Phys. Rev. B}\ }\textbf {\bibinfo {volume} {83}},\ \bibinfo
  {pages} {195139} (\bibinfo {year} {2011})}\BibitemShut {NoStop}%
\bibitem [{\citenamefont {Maciejko}\ and\ \citenamefont
  {Fiete}(2015)}]{maciejko2015fractionalized}%
  \BibitemOpen
  \bibfield  {author} {\bibinfo {author} {\bibfnamefont {J.}~\bibnamefont
  {Maciejko}}\ and\ \bibinfo {author} {\bibfnamefont {G.~A.}\ \bibnamefont
  {Fiete}},\ }\href {https://www.nature.com/articles/nphys3311} {\bibfield
  {journal} {\bibinfo  {journal} {Nature Physics}\ }\textbf {\bibinfo {volume}
  {11}},\ \bibinfo {pages} {385} (\bibinfo {year} {2015})}\BibitemShut
  {NoStop}%
\bibitem [{\citenamefont {Kobayashi}\ \emph {et~al.}(2016)\citenamefont
  {Kobayashi}, \citenamefont {Okumura}, \citenamefont {Yamada}, \citenamefont
  {Machida},\ and\ \citenamefont {Aoki}}]{PhysRevB.94.214501}%
  \BibitemOpen
  \bibfield  {author} {\bibinfo {author} {\bibfnamefont {K.}~\bibnamefont
  {Kobayashi}}, \bibinfo {author} {\bibfnamefont {M.}~\bibnamefont {Okumura}},
  \bibinfo {author} {\bibfnamefont {S.}~\bibnamefont {Yamada}}, \bibinfo
  {author} {\bibfnamefont {M.}~\bibnamefont {Machida}}, \ and\ \bibinfo
  {author} {\bibfnamefont {H.}~\bibnamefont {Aoki}},\ }\href {\doibase
  10.1103/PhysRevB.94.214501} {\bibfield  {journal} {\bibinfo  {journal} {Phys.
  Rev. B}\ }\textbf {\bibinfo {volume} {94}},\ \bibinfo {pages} {214501}
  (\bibinfo {year} {2016})}\BibitemShut {NoStop}%
\bibitem [{\citenamefont {Aoki}(2020)}]{flatsc}%
  \BibitemOpen
  \bibfield  {author} {\bibinfo {author} {\bibfnamefont {H.}~\bibnamefont
  {Aoki}},\ }\href
  {https://link.springer.com/article/10.1007/s10948-020-05474-6} {\bibfield
  {journal} {\bibinfo  {journal} {Journal of Superconductivity and Novel
  Magnetism}\ }\textbf {\bibinfo {volume} {33}},\ \bibinfo {pages} {2341}
  (\bibinfo {year} {2020})}\BibitemShut {NoStop}%
\bibitem [{\citenamefont {Martin}(2020)}]{martin2019moire}%
  \BibitemOpen
  \bibfield  {author} {\bibinfo {author} {\bibfnamefont {I.}~\bibnamefont
  {Martin}},\ }\href {https://doi.org/10.1016/j.aop.2020.168118} {\bibfield
  {journal} {\bibinfo  {journal} {Annals of Physics}\ }\textbf {\bibinfo
  {volume} {417}},\ \bibinfo {pages} {168118} (\bibinfo {year}
  {2020})}\BibitemShut {NoStop}%
\bibitem [{\citenamefont {Panna}\ \emph {et~al.}(2019)\citenamefont {Panna},
  \citenamefont {Landau}, \citenamefont {Gantz}, \citenamefont {Rybak},
  \citenamefont {Tsesses}, \citenamefont {Adler}, \citenamefont {Brodbeck},
  \citenamefont {Schneider}, \citenamefont {H{\"o}fling},\ and\ \citenamefont
  {Hayat}}]{panna2019ultrafast}%
  \BibitemOpen
  \bibfield  {author} {\bibinfo {author} {\bibfnamefont {D.}~\bibnamefont
  {Panna}}, \bibinfo {author} {\bibfnamefont {N.}~\bibnamefont {Landau}},
  \bibinfo {author} {\bibfnamefont {L.}~\bibnamefont {Gantz}}, \bibinfo
  {author} {\bibfnamefont {L.}~\bibnamefont {Rybak}}, \bibinfo {author}
  {\bibfnamefont {S.}~\bibnamefont {Tsesses}}, \bibinfo {author} {\bibfnamefont
  {G.}~\bibnamefont {Adler}}, \bibinfo {author} {\bibfnamefont
  {S.}~\bibnamefont {Brodbeck}}, \bibinfo {author} {\bibfnamefont
  {C.}~\bibnamefont {Schneider}}, \bibinfo {author} {\bibfnamefont
  {S.}~\bibnamefont {H{\"o}fling}}, \ and\ \bibinfo {author} {\bibfnamefont
  {A.}~\bibnamefont {Hayat}},\ }\href
  {https://doi.org/10.1021/acsphotonics.9b00659} {\bibfield  {journal}
  {\bibinfo  {journal} {ACS Photonics}\ }\textbf {\bibinfo {volume} {6}},\
  \bibinfo {pages} {3076} (\bibinfo {year} {2019})}\BibitemShut {NoStop}%
\bibitem [{\citenamefont {Schuetz}\ \emph {et~al.}(2017)\citenamefont
  {Schuetz}, \citenamefont {Kn\"orzer}, \citenamefont {Giedke}, \citenamefont
  {Vandersypen}, \citenamefont {Lukin},\ and\ \citenamefont
  {Cirac}}]{PhysRevX.7.041019}%
  \BibitemOpen
  \bibfield  {author} {\bibinfo {author} {\bibfnamefont {M.~J.~A.}\
  \bibnamefont {Schuetz}}, \bibinfo {author} {\bibfnamefont {J.}~\bibnamefont
  {Kn\"orzer}}, \bibinfo {author} {\bibfnamefont {G.}~\bibnamefont {Giedke}},
  \bibinfo {author} {\bibfnamefont {L.~M.~K.}\ \bibnamefont {Vandersypen}},
  \bibinfo {author} {\bibfnamefont {M.~D.}\ \bibnamefont {Lukin}}, \ and\
  \bibinfo {author} {\bibfnamefont {J.~I.}\ \bibnamefont {Cirac}},\ }\href
  {\doibase 10.1103/PhysRevX.7.041019} {\bibfield  {journal} {\bibinfo
  {journal} {Phys. Rev. X}\ }\textbf {\bibinfo {volume} {7}},\ \bibinfo {pages}
  {041019} (\bibinfo {year} {2017})}\BibitemShut {NoStop}%
\bibitem [{\citenamefont {Min}\ \emph {et~al.}(2008)\citenamefont {Min},
  \citenamefont {Wang}, \citenamefont {Chen}, \citenamefont {Deng},
  \citenamefont {Lu}, \citenamefont {Ming}, \citenamefont {Ning}, \citenamefont
  {Zhou},\ and\ \citenamefont {Yang}}]{Min:08}%
  \BibitemOpen
  \bibfield  {author} {\bibinfo {author} {\bibfnamefont {C.}~\bibnamefont
  {Min}}, \bibinfo {author} {\bibfnamefont {P.}~\bibnamefont {Wang}}, \bibinfo
  {author} {\bibfnamefont {C.}~\bibnamefont {Chen}}, \bibinfo {author}
  {\bibfnamefont {Y.}~\bibnamefont {Deng}}, \bibinfo {author} {\bibfnamefont
  {Y.}~\bibnamefont {Lu}}, \bibinfo {author} {\bibfnamefont {H.}~\bibnamefont
  {Ming}}, \bibinfo {author} {\bibfnamefont {T.}~\bibnamefont {Ning}}, \bibinfo
  {author} {\bibfnamefont {Y.}~\bibnamefont {Zhou}}, \ and\ \bibinfo {author}
  {\bibfnamefont {G.}~\bibnamefont {Yang}},\ }\href {\doibase
  10.1364/OL.33.000869} {\bibfield  {journal} {\bibinfo  {journal} {Opt.
  Lett.}\ }\textbf {\bibinfo {volume} {33}},\ \bibinfo {pages} {869} (\bibinfo
  {year} {2008})}\BibitemShut {NoStop}%
\end{thebibliography}
\end{document}